\documentclass[11pt,a4paper,reqno]{amsart}

\usepackage[margin=0.8in,footskip=0.25in]{geometry}
\usepackage{amsmath, amsthm, amssymb}
\usepackage{graphicx}
\usepackage{booktabs}

\usepackage{pslatex}
\usepackage{array}
\usepackage[square,numbers]{natbib}
\usepackage{amsmath,systeme}
\usepackage{hyperref}

\usepackage[utf8]{inputenc}
\usepackage[T1]{fontenc}

\DeclareUnicodeCharacter{2248}{\ensuremath{\approx}}
\DeclareUnicodeCharacter{2265}{\ensuremath{\geq}}
\DeclareUnicodeCharacter{2273}{\ensuremath{\gtrsim}}
\DeclareUnicodeCharacter{00D7}{\ensuremath{\times}}

\usepackage[nomarkers]{endfloat} 

\date{}

\begin{document}

\title[Topological traps in evolutionary games]
{Topological traps in evolutionary games}

\author{Jose Segovia-Martin$^1$}
\address{$^1$ School of Collective Intelligence (M6 Polytechnic University)}
\address{$^2$ Complex Systems Institute of Paris Ile-de-France (ISCPIF-CNRS)}

\email{Jose.Segovia@um6p.ma}



\begin{abstract}

How cooperation originates and persists among self-interested individuals is a central question in the social and behavioural sciences. In the canonical two-dimensional spatial Prisoner's Dilemma with unconditional imitation introduced by Nowak and May (1992), simulations on a Moore lattice show an abrupt drop in cooperation near the temptation $T\approx5/3$, yet even under these harsh conditions cooperative structures can still arise. However, the nucleation rates of these motifs, and their contribution along the full cooperation curve had not been quantified. Here we show, using large-scale Monte Carlo simulations combined with automatic cluster classification, that on the Moore lattice for $T\ge5/3$ residual cooperation is sustained exclusively by $3\times3$ (or larger) rectangular cooperator bricks, whereas on degree-8 random-regular graphs for $T\gtrsim1.5$ it is dominated by star-like motifs (1 hub + 8 leaves). Once the dynamics becomes nucleation limited, the macroscopic cooperation level is therefore governed by the statistics of a few exceptionally resilient shapes, rather than by many different cooperator motifs. Furthermore, we show that the lattice cooperation collapse near $T=5/3$ is kinetic rather than critical: the reduction in cooperation is not due to a loss of growth capacity of rectangular bricks, but to the progressive destabilisation of the subcritical motifs that dominate just below this threshold. Our results show that residual cooperation at high temptation is a rare-event nucleation phenomenon governed by a small set of topological traps, and highlight the value of motif-level analysis for explaining and engineering cooperation in spatial, social, and technological networks.

\end{abstract}

\maketitle

\section{Main}

The persistence of cooperation among unrelated, self–interested agents is a
long–standing puzzle that permeates the natural and social sciences
\cite{Hamilton1964,Hardin1968,Pennisi2005}.  
Formally, the puzzle is epitomised by the
Prisoner’s Dilemma (PD): although mutual cooperation delivers the highest
collective welfare, unilateral defection strictly dominates and is the only
Nash equilibrium in a well–mixed population
\cite{Axelrod1984,MaynardSmith1982,Nowak2006}.  
Among the many mechanisms proposed to overcome this dilemma
(direct reciprocity, kin selection, punishment, reputation, \emph{inter alia}),
\emph{network reciprocity} (the idea that the pattern of who interacts with
whom matters) remains one of the most influential and best
quantified \cite{Nowak1992,Szabo2007,Roca2009}.

The \emph{Prisoner’s Dilemma} (PD) is the archetypal $2\times2$ game that
formalises the conflict between individual and collective interests.  Each
player chooses either to \emph{cooperate} ($C$) or \emph{defect} ($D$).  If both
cooperate they each receive the \emph{reward} payoff $R$; if both defect they
receive the lower \emph{punishment} payoff $P$.  When one defects while the
other cooperates, the defector earns the highest \emph{temptation} payoff $T$
and the cooperator suffers the lowest \emph{sucker’s} payoff $S$.  The
ordering
\[
T > R > P > S
\]
ensures that defection strictly dominates cooperation: irrespective of the
opponent’s move, $D$ yields a higher individual payoff than $C$.  Consequently
the unique Nash equilibrium is mutual defection $(D,D)$, even though mutual
cooperation $(C,C)$ maximises the collective return $2R$.  The PD thus
encapsulates the core paradox of cooperation and motivates the search for
mechanisms (reciprocity, kin selection, punishment, network structure,
\emph{etc.}) capable of sustaining cooperative behaviour.

Evolutionary game theory provides the mathematical language in which these
mechanisms are typically explored.  In well–mixed (mean-field) settings the
standard description is the \emph{replicator equation}, first derived by
Taylor and Jonker and later reviewed in depth by Hofbauer and Sigmund
\cite{Taylor1978,Hofbauer1998}.  The replicator dynamics link the growth
rate of each strategy to its excess payoff over the population average and
retain the classical Nash–equilibrium concept
\cite{Nash1950} as their fixed points. On graphs, however, individuals
update their strategies through local rules that need not mirror the
replicator flow.  Among the most studied is \emph{unconditional imitation}
(UI): after each round every player adopts the strategy of the neighbour
with the highest payoff, provided that payoff is strictly larger than her
own \cite{Nowak1992}. Despite its extreme simplicity, UI captures an important behavioural
heuristic (i.e.,“copy success”) and, crucially, breaks the symmetry between
cooperators and defectors in spatial PD
\cite{Szabo2007,Roca2009}.  When embedded in a regular lattice this rule
creates positive feedback, where a lone cooperator can recruit neighbouring
defectors, enlarge a compact cluster, and thereby raise its own payoff even
further. The collective outcome is a cooperation plateau that cannot be anticipated from
simple mean-field analysis and that depends sensitively on both the network
structure and the details of the update process.

The seminal lattice study by Nowak and May showed that, under \emph{unconditional
imitation}, spatial clustering of cooperators can protect them from invasion and
generate a striking ``staircase'' dependence of the asymptotic cooperator
fraction on the temptation to defect $T$ \cite{Nowak1992}. In plain terms, their simulations revealed that once a few neighbours begin cooperating they can form a compact “island” that shields its members from being overtaken by defectors. As long as the temptation $T$ is not too large, these cooperative islands survive and even spread, but above certain $T$ thresholds they are finally eroded away.
Subsequent work has generalised those findings to a broad spectrum of update
rules, graph topologies and evolutionary time scales
\cite{Ohtsuki2006,Traulsen2006,Cuesta2015}.  
Yet, more than three decades after the seminal study of Nowak and May, several questions remain open:

\begin{description}
  \item[\textbf{Q1}] \emph{Which concrete cooperator motifs sustain the
        successive cooperation plateaux on different graphs?}

  \item[\textbf{Q2}] \emph{Does the abrupt drop in cooperation at
        intermediate temptation $T$ correspond to a true phase
        transition, or is it a kinetic bottleneck caused by the
        extinction of specific motifs?}

  \item[\textbf{Q3}] \emph{Which motifs account for the residual
        cooperation observed beyond the critical value of~$T$?}
\end{description}

Analytical treatments of the spatial Prisoner’s Dilemma on Nowak and May 2D lattices demonstrate that compact blocks of cooperators forming a rectangular \(k\times\!l\) cluster with \(k,l\ge3\) cannot be invaded even for  $5/3 <T< 8/3$ when $(R,S,T,P)=(1,0,T,0)$ \cite{Szabo2007}. These facts suggest that compact bricks could be
natural candidates to sustain the lattice cooperation plateau, yet their actual nucleation rate
and role in the full cooperation curve have not been quantified. Determining how the availability of these motifs changes across~$T$
is therefore one of the central aims of the present study.

Random--regular (RR) networks with the same degree $k$ offer a natural
counterpoint to the lattice benchmark because they are locally tree-like. In random regular graphs  the concentration of short loops vanishes as $N\to\infty$, so local neighbourhoods 
become increasingly similar to a tree \cite{Szabo2007}. Using RR graphs therefore lets us hold fixed the key ingredients of the Nowak
and May setting (network size, degree, payoff structure, and update rule) while
removing the geometric embedding that supports compact blocks on lattices. We therefore study the same PD dynamics on RR graphs and quantify (i) which cooperator motifs appear, (ii) how often they 
nucleate as the temptation $T$ varies, and (iii) how their growth or extinction
shapes the macroscopic cooperation curve. A key focus is the high temptation
regime. We identify the specific clusters that survive at the largest $T$ and 
test whether residual cooperation there is explained by the rare nucleation of a
small set of exceptionally resilient motifs. 

By comparing the motif census on the lattice and on RR graphs we isolate which
structural features govern 
motif availability, completion, and long-run persistence, thereby linking
microscopic cluster dynamics to macroscopic cooperation across topologies.

\section{Methods}

This section specifies the model and simulation protocol. Section~\ref{subsec:network-structure} describes the network structure, namely either a \(100\times100\) Moore degree-8 lattice (following Nowak and May~\cite{Nowak1992}) or a degree-8 random-regular (RR) graph on \(N = 10^{4}\) nodes. Section~\ref{subsec:strategies-init} details the strategy initialisation, where each node starts as a cooperator or defector with equal probability, the payoff structure, namely a standard Prisoner’s Dilemma matrix with \(R = 1\), \(S = P = 0\) and temptation \(T\in[1,2]\), and the deterministic unconditional imitation update rule, whereby at every synchronous time step each player compares its payoff with those of its neighbours and adopts the strategy of the highest-payoff neighbour whenever that payoff is strictly higher than its own. We scan over \(T\), simulate each configuration for \(100\) time steps, and for every pair (network type, \(T\)) average over \(1000\) independent repetitions differing only in their random initial conditions. Section~\ref{subsec:metrics-clusters} describes the recorded observables: (i) the global cooperation level and the proportions of “consistent-\(C\),’’ “consistent-\(D\),’’ and “oscillator’’ nodes (based on the last two strategies), and (ii) the connected components of the cooperator subgraph, which are automatically labelled \texttt{rectangle}, \texttt{star}, \texttt{chain}, \texttt{loop}, \texttt{compact}, \texttt{cycle}, \texttt{irregular}, etc.\ using simple geometric and graph-theoretic descriptors (bounding-box dimensions, fill ratio, perimeter, interior holes, approximate diameter), together with their temporal classification into stable and unstable clusters.

\subsection{Network structure}
\label{subsec:network-structure}

We consider two network topologies with the same number of nodes and degree.

\paragraph{2D lattice.}
The first topology is a two-dimensional spatial lattice with periodic boundary conditions, forming a Moore neighbourhood for each node. This network consists of \(N = 10{,}000\) nodes arranged in a \(100 \times 100\) grid, where each node is connected to its eight nearest neighbours (four von Neumann neighbours plus four diagonals). The periodic boundary conditions make the lattice topologically equivalent to a torus, avoiding edge effects.

\paragraph{Random-regular network.}
For comparison we consider a homogeneous random network with the same size and degree. Let \(\mathcal{G}_{N,k}\) denote the ensemble of all simple, undirected \(k\)–regular graphs on \(N\) labelled vertices. For \(N=10^4\) and \(k=8\) we draw a single graph \(G=(V,E)\sim \mathrm{Uniform}\bigl(\mathcal{G}_{N,8}\bigr),\) so that every node \(i\in V\) has exactly \(\lvert\mathcal{N}(i)\rvert=8\) neighbors, but without any lattice structure. This random-regular graph is kept fixed while we vary the game parameter \(T\) and run multiple stochastic realisations of the dynamics.

In both topologies the interaction graph is thus simple, undirected and \(8\)–regular on \(N=10^4\) vertices.

\subsection{Strategies and Initialisation}
\label{subsec:strategies-init}

Each node \( i \) in the network adopts a binary strategy, \( s_i \in \{0, 1\} \), where \( 0 \) represents defection and \( 1 \) represents cooperation. At the start of each simulation, strategies are assigned randomly, with each node having an equal probability of adopting either strategy:
\[
P(s_i = 0) = P(s_i = 1) = 0.5.
\]

\subsubsection{Payoff Structure}

The interactions between nodes are governed by the standard Prisoner's Dilemma payoff matrix:
\[
\begin{array}{c|cc}
      & C                      & D                     \\ \hline
C     & (R,R)                  & (S,T)                 \\
D     & (T,S)                  & (P,P)
\end{array}
\]
where \( R = 1 \) (reward), \( S = 0 \) (sucker's payoff), \( P = 0 \) (punishment), and \( T \) (temptation) is varied in the interval \( [1,2] \). In practice we explore \(T\) on the discrete grid
\[
T \in \{1.00, 1.05, \dots, 2.00\} \cup \{1.66, 1.67, 1.68\}.
\]
The total payoff \( \Pi_i \) of node \( i \) is determined by summing the payoffs from pairwise interactions with its neighbours:
\[
\Pi_i = \sum_{j \in \mathcal{N}_i} \pi(s_i, s_j),
\]
where \( \mathcal{N}_i \) is the set of neighbours of node \( i \), and \( \pi(s_i, s_j) \) denotes the payoff obtained by \( i \) when interacting with neighbour \( j \). The payoffs for all possible strategy combinations are:
\[
\pi(1, 1) = R, \quad \pi(1, 0) = S, \quad \pi(0, 1) = T, \quad \pi(0, 0) = P.
\]

\subsubsection{Update Rule: Unconditional Imitation}

The strategy update follows the unconditional imitation rule. At each time step \( t \), each node \( i \) evaluates the total payoffs of itself and its neighbours:
\[
\Pi_j(t), \quad \forall j \in \mathcal{N}_i \cup \{i\}.
\]
Node \( i \) adopts the strategy of the neighbour (or itself) with the highest payoff:
\[
j^* \;=\;\arg\max_{j\in\mathcal{N}(i)\cup\{i\}}\;\Pi_j(t),
\qquad
s_i(t+1) =
\begin{cases}
s_{j^*}(t), & \Pi_{j^*}(t)>\Pi_i(t),\\[6pt]
s_i(t),     & \text{otherwise}.
\end{cases}
\]
If the highest payoff is not strictly greater than its own, the node keeps its strategy.

\subsubsection{Computational Implementation}

The simulations were implemented in Python using the \texttt{NetworkX} library to construct both the lattice and random-regular networks and to manage neighbour relationships. Parallel computation was employed using the \texttt{multiprocessing} library. For each value of \(T\) and each network type, the \(1000\) independent runs were distributed across available CPU cores.

\subsection{Recorded quantities and cluster analysis}
\label{subsec:metrics-clusters}

To characterise the emergent patterns of cooperation, at each time step \(t\) and for each simulation run we record both global and meso-scale descriptors.

\subsubsection{Global cooperation level.}

Let \(s_i(t)\in\{0,1\}\) denote the strategy of node \(i\) at time \(t\), with \(1\equiv C\) (cooperate) and \(0\equiv D\) (defect). The global fraction of cooperators is
\[
f_C(t) \;=\; \frac{1}{N} \sum_{i=1}^N s_i(t).
\]

\subsubsection{Strategy dynamics classification.}

For each node \(i\) and each time \(t\ge 1\) we maintain a rolling history of the two most recent strategies \((s_i(t-1),s_i(t))\). On this basis we classify nodes into three types:

  \textbf{Consistent cooperators}: \(s_i(t-1)=s_i(t)=1\); 

  \textbf{Consistent defectors}: \(s_i(t-1)=s_i(t)=0\);

  \textbf{Oscillators}: \(s_i(t-1)\neq s_i(t)\).

At each time step we record the proportion of nodes in each category, which summarises the stability and volatility of individual strategies.

\subsubsection{Cooperator subgraph and connected components.}

Let \(G=(V,E)\) denote either the lattice or the random-regular graph. The cooperator subgraph at time \(t\) is defined as
\[
G_C(t) \;=\; \bigl(V_C(t),E_C(t)\bigr),
\quad
V_C(t) = \{\,i\in V : s_i(t)=1\,\},
\quad
E_C(t) = \{\,\{u,v\}\in E : u,v\in V_C(t)\,\},
\]
where \(V_C(t)\) is the set of cooperating nodes, and \(E_C(t)\) the set of edges whose endpoints are both cooperators.

We decompose \(G_C(t)\) into its connected components,
\[
\mathcal{K}(t) \;=\; \{\,K^{(\ell)}(t) : \ell = 1,\dots,m_t\,\},
\]
so that \(\mathcal{K}(t)\) is the set of all connected components of \(G_C(t)\). For each component \(K\in\mathcal{K}(t)\) we denote
\[
n_K = |K|, \qquad
e_K = \bigl|\{\,\{u,v\}\in E_C(t) : u,v\in K\,\}\bigr|,
\]
where \(n_K\) is the number of nodes in \(K\) and \(e_K\) the number of edges with both endpoints in \(K\).

We extract the following geometric and topological descriptors:

  \textbf{Perimeter nodes (lattice only).}  
    On the Moore lattice we view clusters on the underlying \(\sqrt{N}\times\sqrt{N}\) grid and use the von Neumann (4–neighbour) structure to define their boundary.  For a cooperator component \(K\subset V_C(t)\), let
    \[
      P_K \;=\;\bigl|\{\,v\in K : \text{fewer than four of its von Neumann neighbours belong to }K\,\}\bigr|
    \]
    be the number of boundary cooperators in \(K\).

  \textbf{Compactness (lattice only).}  
    \[
      \kappa_K = \frac{n_K}{P_K}\quad(\text{higher }\kappa_K\text{ means denser, more compact clusters}).
    \]

  \textbf{Bounding box and fill ratio (lattice only).}  
    On the lattice we consider the minimal bounding box of \(K\) aligned with the
lattice axes, with width \(w_K\), height \(h_K\) and area \(A_K = w_K h_K\).  The fill ratio is
    \[
      \phi_K = \frac{n_K}{A_K},
    \]
    and we use \(\min(w_K,h_K)\) and the elongation \(\max(w_K,h_K)/\min(w_K,h_K)\) to detect thin or highly elongated clusters.

  \textbf{Interior holes (lattice only).}  
    Within the bounding box of \(K\) we consider the complement of \(K\) and count the finite connected components of this complement. The resulting number of interior holes is denoted \(H_K\).

  \textbf{Diameter and aspect ratio (random-regular graph only).} 
    For components in random-regular networks, we approximate the graph-theoretic diameter
    \[
      \delta_K = \max_{u,v\in K}\,\mathrm{dist}_{G_C}(u,v),
    \]
    and define an aspect ratio
    \[
      \alpha_K = \frac{\delta_K}{\max(n_K-1,1)}\quad(\text{chain‐like if }\alpha_K\text{ is large}).
    \]

We then classify each cluster \(K\) according to its descriptors.  In particular:
\[
\text{type}(K)=
\begin{cases}
\texttt{single\_node}, & n_K=1,\\[4pt]
\texttt{star\_}k+L, & \text{\(K\) is a tree with no degree-2 nodes and \(k\) high-degree centres with }L\text{ leaves},\\[4pt]
\texttt{large\_cluster}, & n_K\ge 100,\\
\end{cases}
\]
and otherwise we proceed by topology.

On the 2D lattice we use the bounding box, fill ratio, holes and compactness:
\[
\text{type}(K)=
\begin{cases}
\texttt{rectangle\_}w_K\texttt{x}h_K, &
    \phi_K = 1\ \text{and every site in the bounding box belongs to }K,\\[4pt]
\texttt{chain\_}n_K, &
    \min(w_K,h_K)=1\ \text{or}\ \bigl(\tfrac{\max(w_K,h_K)}{\min(w_K,h_K)}\ge 3
    \ \text{and}\ \phi_K\ge 0.6\bigr),\\[4pt]
\texttt{loop\_}n_K, &
    H_K \ge 1,\\[4pt]
\texttt{compact\_}n_K, &
    \kappa_K > 1.5,\\[4pt]
\texttt{irregular\_}n_K, & \text{otherwise.}
\end{cases}
\]

On random-regular networks there is no geometric embedding, so we rely on edge count and diameter:
\[
\text{type}(K)=
\begin{cases}
\texttt{cycle\_}n_K, & e_K \ge n_K,\\[4pt]
\texttt{chain\_}n_K, & e_K < n_K\ \text{and}\ \alpha_K > 0.7,\\[4pt]
\texttt{irregular\_}n_K, & \text{otherwise.}
\end{cases}
\]

In practice this classification is implemented exactly as in
\texttt{classify\_cluster\_shape()} of our code, yielding human‐readable labels such as \texttt{"rectangle\_4x5"}, \texttt{"star\_2+6"}, \texttt{"compact\_12"} or \texttt{"chain\_5"}.  These labels feed into our statistical analysis of cooperator‐cluster morphologies over time.

\subsubsection{Temporal stability of clusters.}

To distinguish persistent patterns from transient fluctuations, we compute the overlap between clusters at times \(t-1\) and \(t\). A cluster at time \(t\) is labeled \emph{stable} if its node set matches a cluster at time \(t-1\); otherwise, it is labeled \emph{unstable}. We record the number of stable \(N_s\) and unstable \(N_u\) clusters at each step, and aggregate their type distributions to study persistence and volatility in cooperative organisation.

\subsubsection{Data aggregation and visualisation.}

For each network type and temptation value \(T\), we summarise the simulation output at the final time step \(t=t_{\max}=100\) by aggregating over the \(1000\) independent repetitions.

First, we compute the mean and standard deviation (across repetitions) of:
\begin{itemize}
  \item the global cooperation level \(f_C(t_{\max})\);
  \item the fractions of each strategy type at \(t_{\max}\) (consistent‐\(C\), consistent‐\(D\), oscillators);
  \item the numbers of stable and unstable cooperator clusters at \(t_{\max}\).
\end{itemize}
These quantities are plotted as functions of \(T\), with separate curves for each network type and shaded bands representing \(\pm 1\) standard deviation across repetitions. This yields summary profiles of cooperation, strategy
composition, and cluster counts for each topology.

Second, for each network type, temptation value \(T\) and cluster shape label (motif) \(m\), we compute the share of cooperator clusters of type \(m\) at the final time:
\[
\mathrm{share}(m \mid T)
\;=\;
\frac{N_m(T)}{N_{\mathrm{tot}}(T)},
\]
where both numerator and denominator are summed over all repetitions. We do this separately for stable and unstable clusters. The resulting motif compositions are shown as heatmaps with motif types on the vertical axis, \(T\) on the horizontal axis, and colour indicating \(\text{share}(m\,|\,T,\text{network})\), allowing us to compare how the distribution of cluster morphologies changes across temptation regimes and between topologies.

\section{Results}

\subsection{General comparison of cooperation profiles and clusters: 2D lattice vs. Random-Regular graph}

\par

\begin{figure}[htbp]
  \centering
  \includegraphics[width=0.9\linewidth]{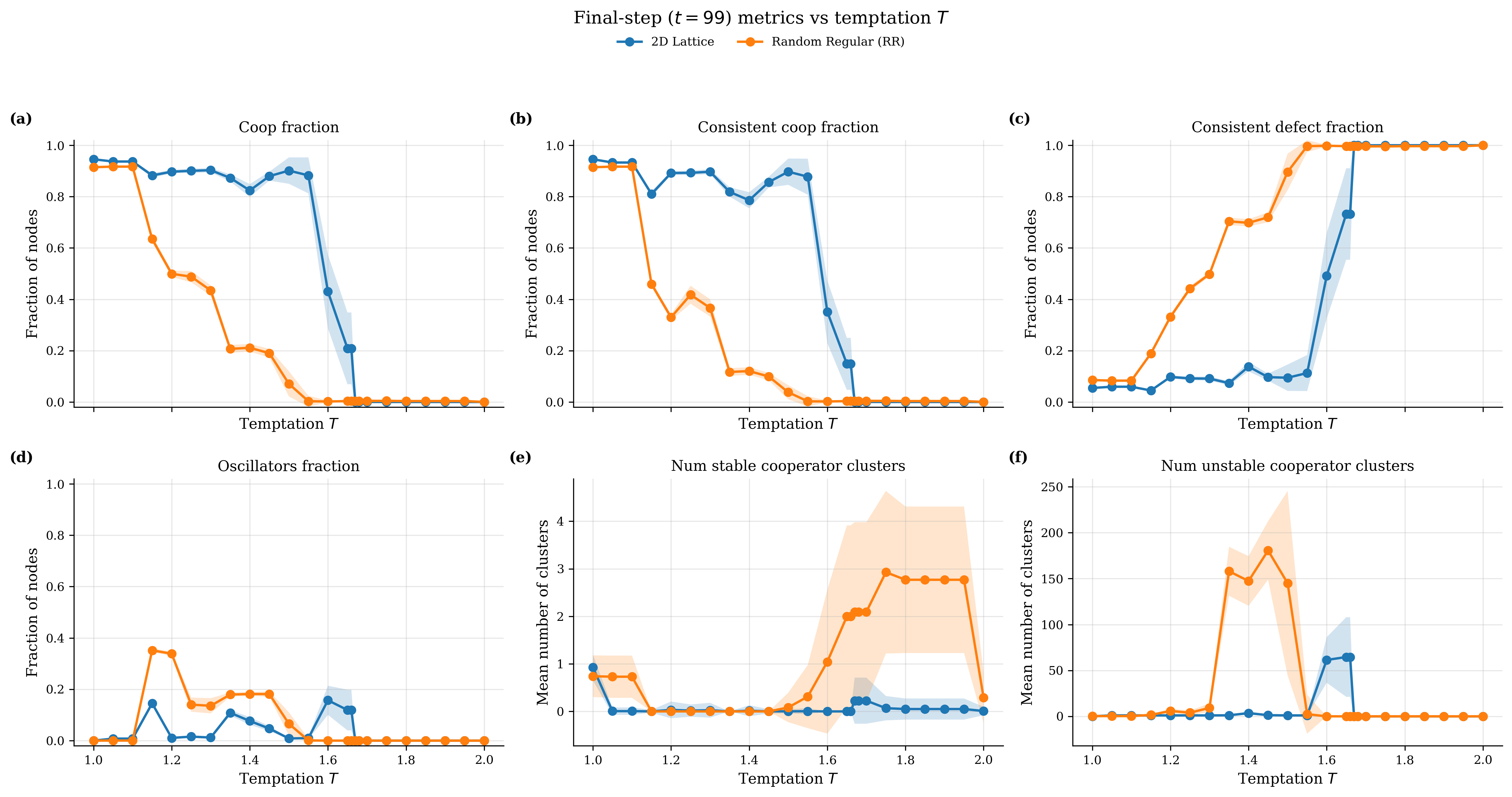}
  \caption{Final-step (\(t = 99\)) network-level metrics as a function of temptation \(T\) for 2D lattices and random-regular graphs. Lines show means across repetitions; shaded regions indicate \(\pm 1\) standard deviation.}
  \label{fig:metrics-vs-T}
\end{figure}

Figure~\ref{fig:metrics-vs-T} shows that the global cooperation curves
$f_C(T)$ differ qualitatively across the two topologies even though size, degree
and update rule are held fixed. On the $100\times100$ Moore lattice,
cooperation remains high for most of $T\in[1,1.55]$ and then collapses sharply
as $T$ approaches the geometric threshold $T_g=5/3$, dropping from
$f_C\simeq0.43$ at $T=1.60$ to $f_C\simeq0.21$ at $T=1.65$ and to
$f_C\approx2\times10^{-4}$ at $T\approx1.67$ (Table~\ref{tab:lattice-clusters}).
By contrast, on degree--8 random--regular graphs the decay begins much earlier:
$f_C$ is close to $0.92$ only up to $T\simeq1.10$, then decreases steadily for
$1.10<T<1.50$ (e.g., down to $f_C\simeq0.19$ at $T=1.45$) and enters a rare--event plateau with $f_C=\mathcal{O}(10^{-3})$ for
$1.55\le T\le1.95$ (Table~\ref{tab:rr-clusters}). The accompanying consistency
metrics in Fig.~\ref{fig:metrics-vs-T} show the same contrast. The RR graph
quickly becomes dominated by consistent defectors as $T$ increases, whereas the
lattice maintains a substantial cooperation until very near
$T_g$.

\begin{figure}[htbp]
  \centering
  \includegraphics[width=0.9\linewidth]{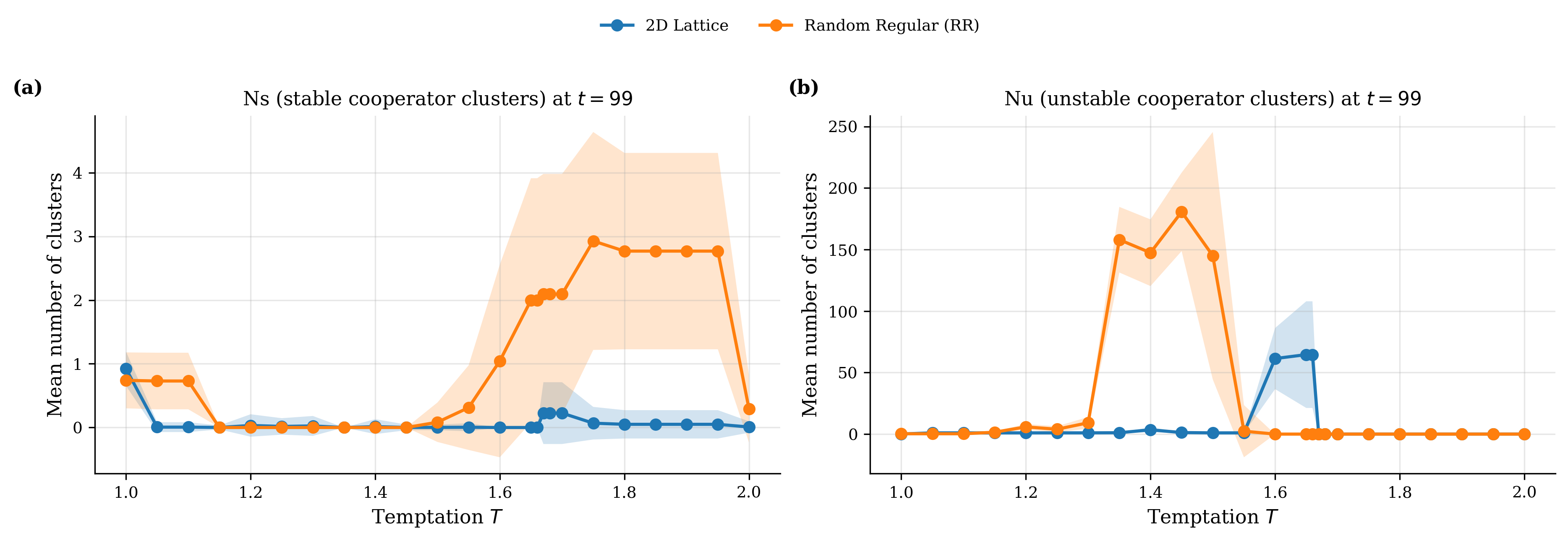}
  \caption{Mean number of stable (\(N_s\)) and unstable (\(N_u\)) cooperator clusters at \(t = 99\) as a function of temptation \(T\), for 2D lattices and random-regular graphs. Shaded regions indicate \(\pm 1\) standard deviation.}
  \label{fig:Ns-Nu-vs-T}
\end{figure}

Figure~\ref{fig:Ns-Nu-vs-T} and Tables~\ref{tab:lattice-clusters}--\ref{tab:rr-clusters}
reveal that these macroscopic curves are reached through very different
mesoscopic cluster organisation. On the lattice, low and moderate temptation
typically lead to a macroscopic cooperative component (hence small
$\langle N_u\rangle$ and, in the easiest cases, a stable component count near
one), whereas the approach to $T_g$ is marked by a surge of unstable clusters. For $T\simeq1.60$--$1.66$ we observe $\langle N_u\rangle\approx60$--$65$
with essentially no stable clusters, consistent with a nucleation-limited
slowdown rather than an immediate loss of brick viability. On the random--regular
graph the intermediate regime $1.10<T<1.50$ is instead a fragmentation regime, where
$\langle N_s\rangle$ collapses to zero and $\langle N_u\rangle$ grows by roughly
two orders of magnitude, peaking around $\langle N_u\rangle\approx180$ at
$T=1.45$ (Table~\ref{tab:rr-clusters}). Only beyond $T\gtrsim1.5$ does
$\langle N_u\rangle$ fall back toward zero as the system settles into a small
number of stable residual components ($\langle N_s\rangle\approx2$--$3$ for much
of $1.6\le T\le1.9$).

\begin{figure}[htbp]
  \centering
  \includegraphics[width=0.9\linewidth]{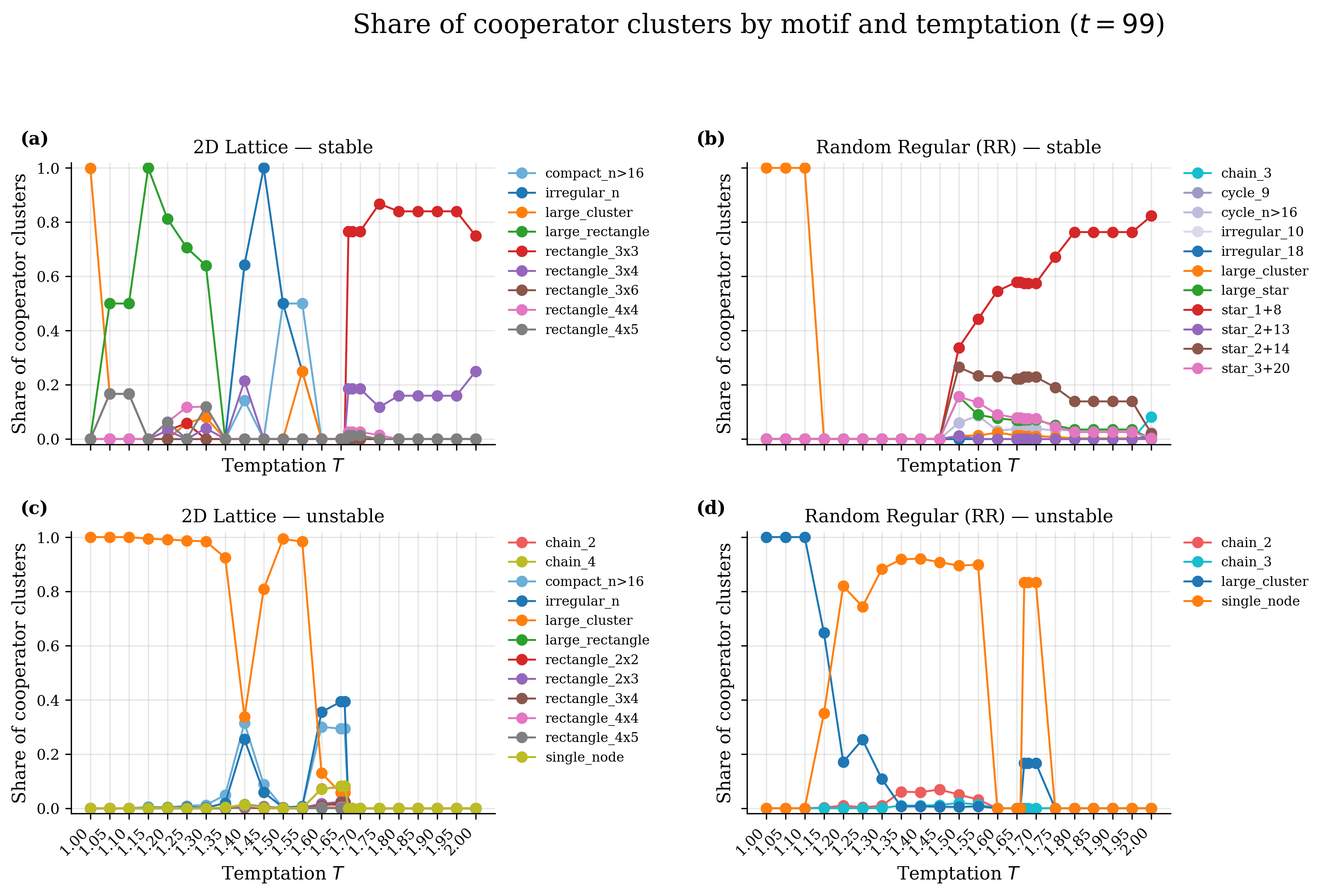}
  \caption{Share of stable and unstable clusters by motif type and temptation at \(t = 99\) for 2D lattices and Random-Regular graphs.}
  \label{fig:motif-lines}
\end{figure}

The motif census in Fig.~\ref{fig:motif-lines} (see also the shares of dominant
stable shapes in Tables~\ref{tab:lattice-clusters} and~\ref{tab:rr-clusters})
links these aggregate profiles to a small set of surviving motifs, which
differs between the two networks. At high $T$ on the lattice, stable
cooperation is carried exclusively by fully cooperative rectangular bricks
($3\times3$ or larger). On the random--regular graph, stable motifs at high $T$
are instead almost entirely completed stars, primarily \texttt{star\_1+8} and
\texttt{star\_2+14}. Thus, even at fixed degree $k=8$, removing geometric
embedding shifts the dominant resilient cooperative building blocks from compact
rectangles to star-like traps. In both topologies, however, at extreme
temptation the most frequent stable finite cooperator motif contains a
cooperator vertex with $n=k$ cooperative neighbours.

\subsection{Two–dimensional square lattice (\(k=8\))}
\label{sec:results:lattice}

\subsubsection{Local payoff landscape of a \(3\times3\) brick}
\label{sec:results:lattice:matrix}

Building on the local payoff analysis of rectangular domains in Szabó and Fáth~\cite{Szabo2007}, we re-derive the payoffs for a solitary \(3\times3\) cooperative brick.

Let a solitary \(3\times3\) block of cooperators be embedded in
defectors. Using
Moore neighbourhood (\(k=8\)) with Prisoner’s Dilemma pay-offs
\((R,S,T,P)=(1,0,T,0)\), the pay-off of every site in the \(7\times7\) window
centred on the brick is

\[
\small
\begin{array}{ccccccc}
0 & 0 & 0 & 0   & 0   & 0 & 0\\
0 & T & 2T & 3T & 2T & T & 0\\
0 & 2T & 3 & 5  & 3   & 2T & 0\\
0 & 3T & 5 & 8  & 5   & 3T & 0\\
0 & 2T & 3 & 5  & 3   & 2T & 0\\
0 & T  & 2T & 3T & 2T & T & 0\\
0 & 0 & 0 & 0   & 0   & 0 & 0
\end{array}
\tag{$1$}
\]

\noindent
We label this array as \((1)\).  Reading the four distinctive entries of
\((1)\) we find

Interior, side-border, and corner cooperators therefore earn, respectively,

\[
\pi^{\text{int}}_{C}=8,\qquad
\pi^{\text{side}}_{C}=5,\qquad
\pi^{\text{corner}}_{C}=3,
\]

while the most fortunate defector surrounding the brick gets

\[
\pi^{\max}_{D}=3T .
\]

\subsubsection{Growth, marginal stability and decay}
\label{sec:results:lattice:thresholds}

Under unconditional imitation a node copies a neighbour only when that
neighbour’s pay-off is strictly larger.
Inspection of the inequalities implied by (\(1\)) yields three dynamical
regimes:

\begin{center}
\begin{tabular}{@{}lll@{}}
Condition on \(T\) & Dominant inequality           & Regime\\
\(T<5/3\)           & \(\pi_C^{\text{side}}>\pi_D^{\max}\) & growth\\
\(5/3\le T<8/3\)   & \(\pi_D^{\max}<\pi_C^{\text{int}}\) and
                     \(\pi_D^{\max}\ge\pi_C^{\text{side}}\) & stability\\
\(T\ge8/3\)        & \(\pi_D^{\max}>\pi_C^{\text{int}}\)  & decay\\
\end{tabular}
\end{center}

Because the numerical simulations are restricted to \(T\le2\), only the first two
regimes are observed in the data.

\subsubsection{Numerical census and kinetic interpretation}

Table~\ref{tab:lattice-clusters} presents summary statistics for the lattice, including the fraction of cooperation, the numbers of stable and unstable cooperator clusters, and the share of stable clusters.

We define the rim-growth velocity as \(v(T) \equiv 5 - 3T\), which quantifies the effective rate at which the boundary of a cooperative cluster advances in the lattice. This expression is derived under the simplifying assumption that the growing shape is an isolated \(3\times3\) cooperator brick in a sea of defectors. While this velocity estimate is exact for such minimal brick shapes at the invasion threshold, it serves as an approximation in practice, since sub-critical shapes may exhibit slower growth.

\begin{figure}[ht]
  \centering
  \includegraphics[width=0.8\textwidth]{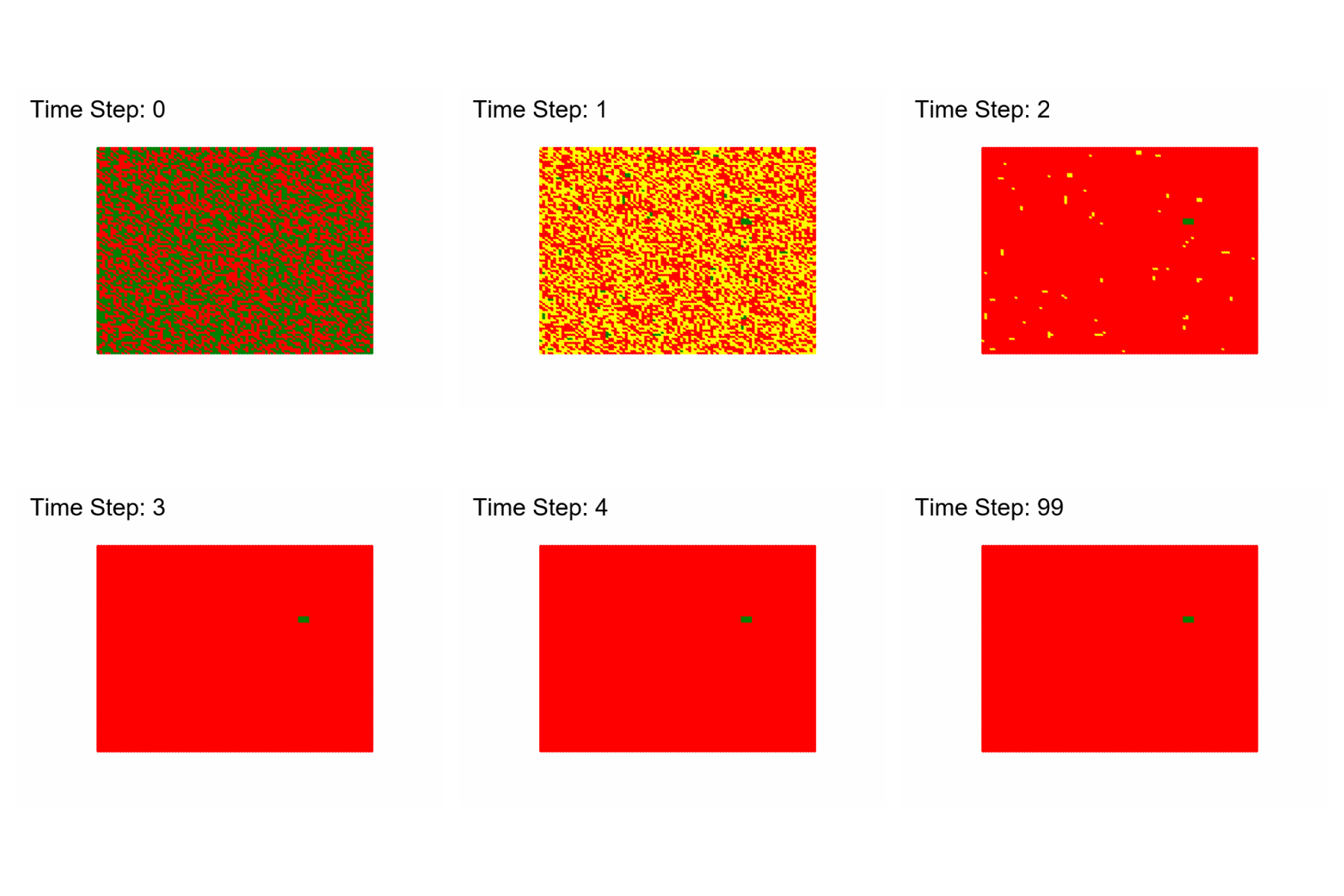}
  \caption{Time evolution of cooperator clusters on a $100\times100$ lattice for $T=1.7$. Snapshots at time steps $t=0,1,2,3,4,99$ show an abrupt halt to all growth. Only 3x3 (or larger bricks) are stable.}
  \label{fig:T1.7}
\end{figure}

\begin{itemize}
\item[(a)]
  \textbf{\(T<1.60\).}
  Most cooperators have coalesced into
  \texttt{large\_clusters}, giving
  \(\bar f_{C}\sim0.945\). Even though \(T = 1.0 < T_g = 5/3\) implies positive rim-growth velocity \(v(T) = 5 - 3T = 2\), which should allow cooperative clusters to expand, cooperation fails to reach 100\% due to the formation of defecting strips that never disappear. These ``defector filaments'' are 1D chains of defectors that can remain stable (see Fig.~\ref{fig:T1.0}). As \(T\) increases to values like \(T = 1.4\), these filaments thicken and percolate through the system, creating rugged frontiers between cooperator blocks (see Fig.~\ref{fig:T1.4}). These patterns prevent cooperator clusters from merging, so the cooperation level stays below one even when the rim growth velocity is positive.

\begin{figure}[ht]
  \centering
  \includegraphics[width=0.8\textwidth]{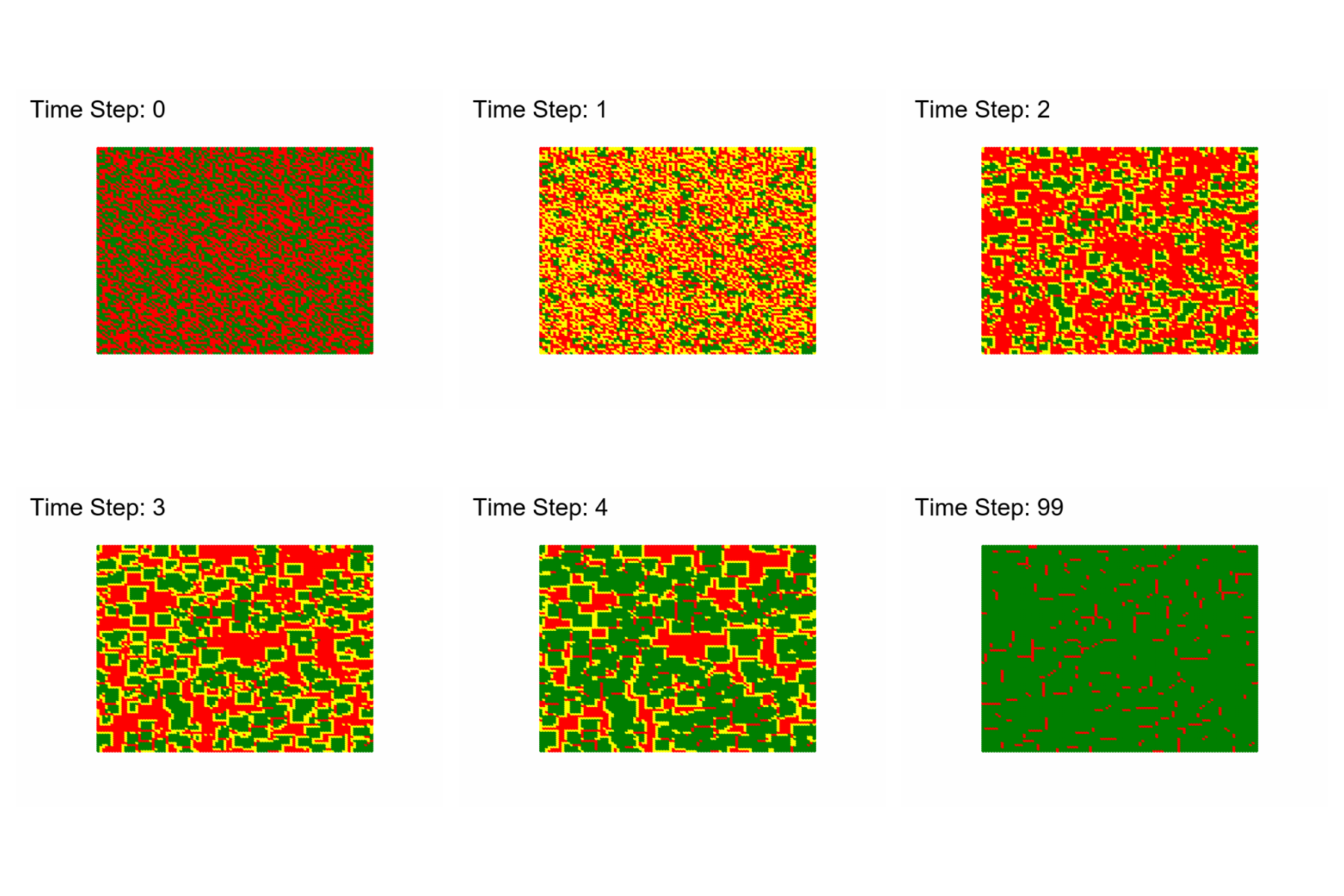}
  \caption{Time evolution of cooperator clusters on a $100\times100$ lattice for $T=1.0$. Snapshots at time steps $t=0,1,2,3,4,99$ show the expansion of cooperative motifs and the persistence of defecting filaments.}
  \label{fig:T1.0}
\end{figure}

\begin{figure}[ht]
  \centering
  \includegraphics[width=0.8\textwidth]{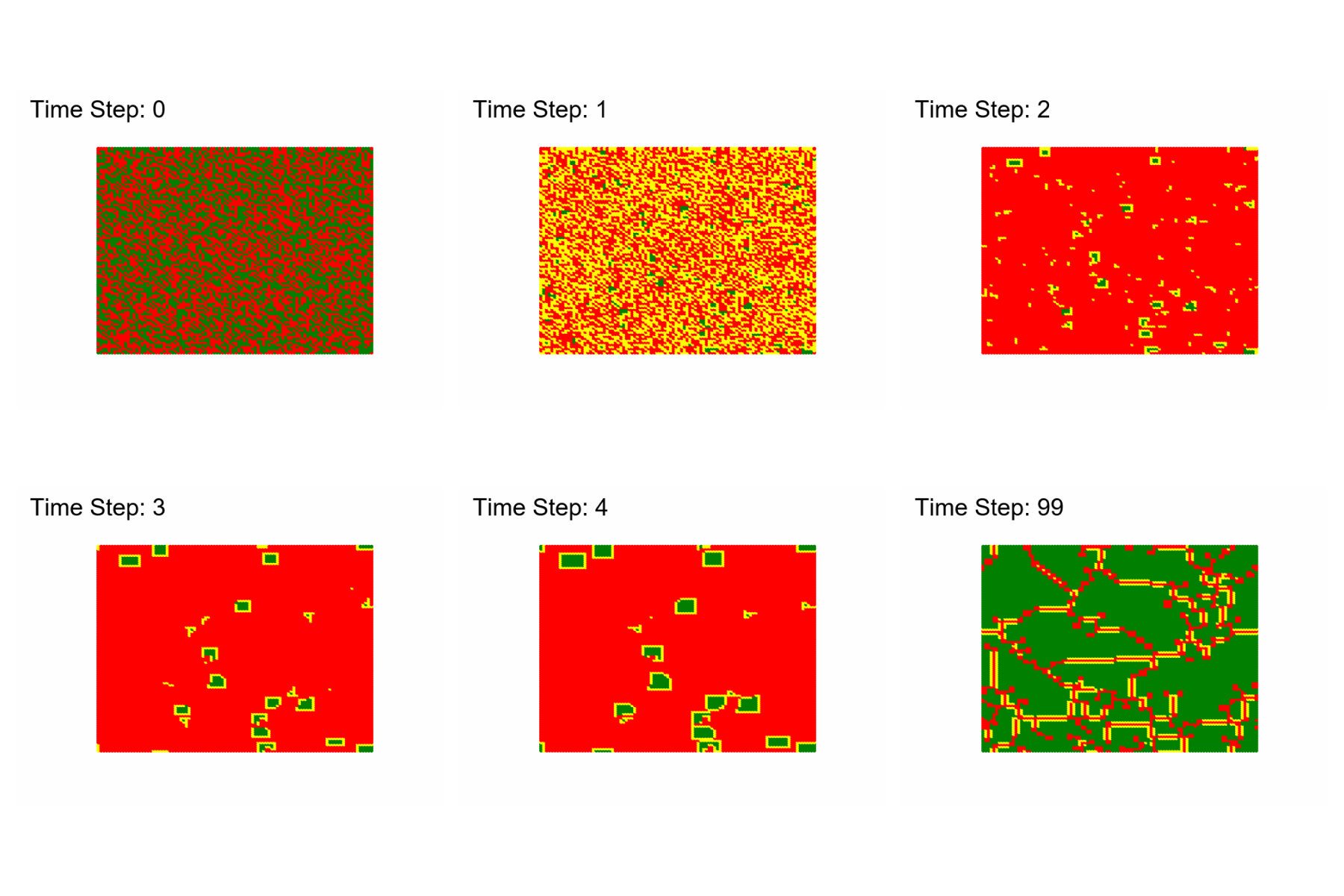}
  \caption{Time evolution of cooperator clusters on a $100\times100$ lattice for $T=1.4$. Snapshots at time steps $t=0,1,2,3,4,99$ illustrate the expansion of cooperative bricks and other irregular shapes and the stability of certain defector structures at $t=99$.}
  \label{fig:T1.4}
\end{figure}

\item[(b)]
  \textbf{\(1.60<T<1.67\).}
  The velocity shrinks to \(0.2\).
  Consequently each snapshot is still populated by dozens of
  growing but unfinished domains
  (\(\langle N_{\mathrm u}\rangle\approx65\) at \(T=1.66\)), and $\bar f_{C} \sim 0.21$.
  The reduction is not due to a loss of growth capacity of the
\(3\times3\) or \(3\times4\) rectangles (both still satisfy
\(3T<5\)).
Rather, for \(T\gtrsim1.60\) all competing cooperator shapes—
large, compact, irregular, or chain–like—either become unstable
or stop expanding.
  These sub-critical shapes (i.e., compact and irregular motifs) are the dominant nucleation sites at
  \(T<1.60\), and once their growth halts the effective seed density falls
  by roughly two orders of magnitude, preventing percolation within
  \(t=99\). See Fig.~\ref{fig:T1.6}.

\begin{figure}[ht]
  \centering
  \includegraphics[width=0.8\textwidth]{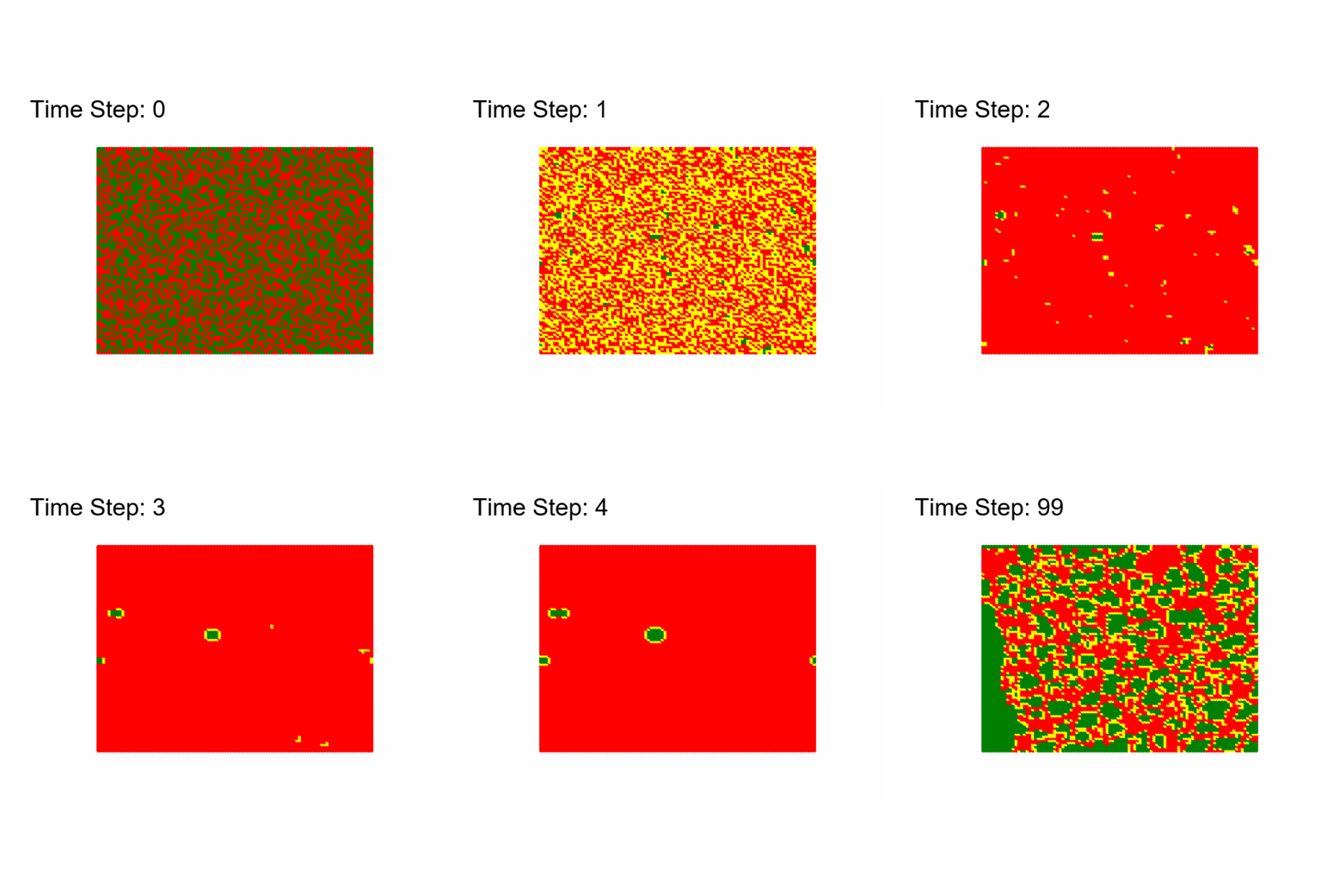}
  \caption{Time evolution of cooperator clusters on a $100\times100$ lattice for $T=1.6$. Snapshots at time steps $t=0,1,2,3,4,99$ show the limited expansion of sub-critical cooperative shapes and the persistence of defecting shapes at $t=99$.}
  \label{fig:T1.6}
\end{figure}

\item[(c)]
  \textbf{\(T=5/3\approx1.67\).}
  Now \(v=0\).
  Growth stops abruptly. Any cluster that is not
  a fully-saturated \(3\times3\) (or larger rectangular) brick
  is erased in the very next timestep.
  Only those runs that had nucleated at least one brick before
  \(t=98\) retain cooperation, yielding
  \(\bar f_{C}\approx2.2\times10^{-4}\).

\item[(d)] 
\textbf{\(1.67<T\le2.0\).}. Bricks \(3\times3\) (or any larger compact
rectangle) are neutrally stable (i.e.\ neither growing nor shrinking) until
\(T>8/3\) provided the first layer of surrounding defectors in its immediate
rim are not boosted by stray cooperators (which is likely to happen in the
first timesteps). As \(T\) increases towards \(2\), the payoff advantage of a
boosted defector increases, which can explain the slight decline of
\(\bar f_{C}(T)\) reported in Table~\ref{tab:lattice-clusters}. See Fig.~\ref{fig:T1.7}.

\end{itemize}

\begin{table}[t]
\centering
\caption{Square--lattice cluster statistics at $t=99$. Results are averaged over $R=1000$ independent runs. One--sigma run-to-run fluctuations are reported for $f_C$. $N_s$ stands for the mean number of stable cooperator clusters at $t=99$. $N_u$ stands for the mean number of unstable cooperator clusters at $t=99$. The last column lists the two most frequent stable shapes and their relative shares of cooperator clusters.}
\label{tab:lattice-clusters}
\begin{tabular}{lcccc}
\toprule
$T$ & $f_C \pm \sigma$ & $\langle N_s \rangle \pm \sigma$ & $\langle N_u \rangle \pm \sigma$ & Dominant stable shapes (share) \\
\midrule
$1.00$ & $0.945 \pm 0.004$ & $0.924 \pm 0.269$ & $0.077 \pm 0.267$ & large\_cluster (0.999), large\_rectangle (0.001) \\
$1.05$ & $0.937 \pm 0.004$ & $0.006 \pm 0.077$ & $0.999 \pm 0.032$ & large\_rectangle (0.500), large\_cluster (0.167) \\
$1.10$ & $0.937 \pm 0.004$ & $0.006 \pm 0.077$ & $0.999 \pm 0.032$ & large\_rectangle (0.500), large\_cluster (0.167) \\
$1.15$ & $0.882 \pm 0.009$ & $0.001 \pm 0.032$ & $1.006 \pm 0.077$ & large\_rectangle (1.000) \\
$1.20$ & $0.897 \pm 0.007$ & $0.032 \pm 0.176$ & $1.012 \pm 0.109$ & large\_rectangle (0.812), rectangle\_4x4 (0.062) \\
$1.25$ & $0.900 \pm 0.008$ & $0.017 \pm 0.129$ & $1.014 \pm 0.118$ & large\_rectangle (0.706), rectangle\_4x4 (0.118) \\
$1.30$ & $0.903 \pm 0.009$ & $0.025 \pm 0.156$ & $1.018 \pm 0.133$ & large\_rectangle (0.640), rectangle\_4x4 (0.120) \\
$1.35$ & $0.873 \pm 0.014$ & $0.000 \pm 0.000$ & $1.111 \pm 0.356$ & \textemdash \\
$1.40$ & $0.824 \pm 0.025$ & $0.014 \pm 0.118$ & $3.501 \pm 2.002$ & irregular\_n (0.643), rectangle\_3x4 (0.214) \\
$1.45$ & $0.880 \pm 0.017$ & $0.002 \pm 0.045$ & $1.248 \pm 0.579$ & irregular\_n (1.000) \\
$1.50$ & $0.901 \pm 0.051$ & $0.002 \pm 0.045$ & $1.004 \pm 0.100$ & compact\_n$>$16 (0.500), irregular\_n (0.500) \\
$1.55$ & $0.883 \pm 0.070$ & $0.004 \pm 0.063$ & $1.013 \pm 0.144$ & compact\_n$>$16 (0.500), irregular\_n (0.250) \\
$1.60$ & $0.430 \pm 0.143$ & $0.000 \pm 0.000$ & $61.404 \pm 24.832$ & \textemdash \\
$1.65$ & $0.209 \pm 0.140$ & $0.000 \pm 0.000$ & $64.600 \pm 43.345$ & \textemdash \\
$1.66$ & $0.209 \pm 0.140$ & $0.000 \pm 0.000$ & $64.600 \pm 43.345$ & \textemdash \\
1.67 & $2.2\times 10^{-4} \pm 4.9\times 10^{-4}$ & $0.226 \pm 0.485$ & $0.000 \pm 0.000$ & rectangle\_3x3 (0.765), rectangle\_3x4 (0.186) \\
1.68 & $2.2\times 10^{-4} \pm 4.9\times 10^{-4}$ & $0.226 \pm 0.485$ & $0.000 \pm 0.000$ & rectangle\_3x3 (0.765), rectangle\_3x4 (0.186) \\
1.70 & $2.2\times 10^{-4} \pm 4.9\times 10^{-4}$ & $0.226 \pm 0.485$ & $0.000 \pm 0.000$ & rectangle\_3x3 (0.765), rectangle\_3x4 (0.186) \\
1.75 & $6.4\times 10^{-5} \pm 2.4\times 10^{-4}$ & $0.068 \pm 0.256$ & $0.000 \pm 0.000$ & rectangle\_3x3 (0.868), rectangle\_3x4 (0.118) \\
1.80 & $4.7\times 10^{-5} \pm 2.1\times 10^{-4}$ & $0.050 \pm 0.223$ & $0.000 \pm 0.000$ & rectangle\_3x3 (0.840), rectangle\_3x4 (0.160) \\
1.85 & $4.7\times 10^{-5} \pm 2.1\times 10^{-4}$ & $0.050 \pm 0.223$ & $0.000 \pm 0.000$ & rectangle\_3x3 (0.840), rectangle\_3x4 (0.160) \\
1.90 & $4.7\times 10^{-5} \pm 2.1\times 10^{-4}$ & $0.050 \pm 0.223$ & $0.000 \pm 0.000$ & rectangle\_3x3 (0.840), rectangle\_3x4 (0.160) \\
1.95 & $4.7\times 10^{-5} \pm 2.1\times 10^{-4}$ & $0.050 \pm 0.223$ & $0.000 \pm 0.000$ & rectangle\_3x3 (0.840), rectangle\_3x4 (0.160) \\
2.00 & $7.8\times 10^{-6} \pm 8.8\times 10^{-5}$ & $0.008 \pm 0.089$ & $0.000 \pm 0.000$ & rectangle\_3x3 (0.750), rectangle\_3x4 (0.250) \\
\bottomrule
\end{tabular}
\end{table}

\begin{table}[t]
\centering
\caption{Random--regular cluster statistics at $t=99$. Results are averaged over $R=1000$ independent runs. One--sigma run-to-run fluctuations are reported for $f_C$. $N_s$ stands for the mean number of stable cooperator clusters at $t=99$. $N_u$ stands for the mean number of unstable cooperator clusters at $t=99$. The last column lists the two most frequent stable shapes and their relative shares of cooperator clusters.}
\label{tab:rr-clusters}
\begin{tabular}{lcccc}
\toprule
$T$ & $f_C \pm \sigma$ & $\langle N_s \rangle \pm \sigma$ & $\langle N_u \rangle \pm \sigma$ & Dominant stable shapes (share) \\
\midrule
$1.00$ & $0.914 \pm 0.003$ & $0.740 \pm 0.439$ & $0.260 \pm 0.439$ & large\_cluster (1.000) \\
$1.05$ & $0.917 \pm 0.003$ & $0.730 \pm 0.444$ & $0.270 \pm 0.444$ & large\_cluster (1.000) \\
$1.10$ & $0.917 \pm 0.003$ & $0.730 \pm 0.444$ & $0.270 \pm 0.444$ & large\_cluster (1.000) \\
$1.15$ & $0.635 \pm 0.011$ & $0.000 \pm 0.000$ & $1.546 \pm 0.727$ & \textemdash \\
$1.20$ & $0.499 \pm 0.012$ & $0.000 \pm 0.000$ & $5.858 \pm 2.377$ & \textemdash \\
$1.25$ & $0.488 \pm 0.022$ & $0.000 \pm 0.000$ & $3.952 \pm 2.402$ & \textemdash \\
$1.30$ & $0.434 \pm 0.020$ & $0.000 \pm 0.000$ & $9.199 \pm 4.739$ & \textemdash \\
$1.35$ & $0.207 \pm 0.015$ & $0.000 \pm 0.000$ & $158.037 \pm 26.669$ & \textemdash \\
$1.40$ & $0.211 \pm 0.015$ & $0.000 \pm 0.000$ & $147.425 \pm 27.103$ & \textemdash \\
$1.45$ & $0.190 \pm 0.017$ & $0.000 \pm 0.000$ & $180.654 \pm 31.886$ & \textemdash \\
$1.50$ & $0.071 \pm 0.049$ & $0.083 \pm 0.310$ & $145.008 \pm 100.554$ & star\_1+8 (0.337), star\_2+14 (0.265) \\
$1.55$ & $0.003 \pm 0.021$ & $0.312 \pm 0.664$ & $2.377 \pm 21.226$ & star\_1+8 (0.442), star\_2+14 (0.234) \\
$1.60$ & $0.002 \pm 0.004$ & $1.044 \pm 1.511$ & $0.000 \pm 0.000$ & star\_1+8 (0.546), star\_2+14 (0.231) \\
$1.65$ & $0.003 \pm 0.004$ & $1.997 \pm 1.916$ & $0.000 \pm 0.000$ & star\_1+8 (0.579), star\_2+14 (0.222) \\
$1.66$ & $0.003 \pm 0.004$ & $1.997 \pm 1.916$ & $0.000 \pm 0.000$ & star\_1+8 (0.579), star\_2+14 (0.222) \\
$1.67$ & $0.004 \pm 0.008$ & $2.094 \pm 1.887$ & $0.006 \pm 0.190$ & star\_1+8 (0.575), star\_2+14 (0.229) \\
$1.68$ & $0.004 \pm 0.008$ & $2.094 \pm 1.887$ & $0.006 \pm 0.190$ & star\_1+8 (0.575), star\_2+14 (0.229) \\
$1.70$ & $0.004 \pm 0.008$ & $2.094 \pm 1.887$ & $0.006 \pm 0.190$ & star\_1+8 (0.575), star\_2+14 (0.229) \\
$1.75$ & $0.004 \pm 0.003$ & $2.928 \pm 1.710$ & $0.000 \pm 0.000$ & star\_1+8 (0.672), star\_2+14 (0.191) \\
$1.80$ & $0.004 \pm 0.003$ & $2.769 \pm 1.541$ & $0.000 \pm 0.000$ & star\_1+8 (0.763), star\_2+14 (0.139) \\
$1.85$ & $0.004 \pm 0.003$ & $2.769 \pm 1.541$ & $0.000 \pm 0.000$ & star\_1+8 (0.763), star\_2+14 (0.139) \\
$1.90$ & $0.004 \pm 0.003$ & $2.769 \pm 1.541$ & $0.000 \pm 0.000$ & star\_1+8 (0.763), star\_2+14 (0.139) \\
$1.95$ & $0.004 \pm 0.003$ & $2.769 \pm 1.541$ & $0.000 \pm 0.000$ & star\_1+8 (0.763), star\_2+14 (0.139) \\
2.00 & $2.6\times 10^{-4} \pm 5.0\times 10^{-4}$ & $0.293 \pm 0.530$ & $0.000 \pm 0.000$ & star\_1+8 (0.823), chain\_3 (0.082) \\
\bottomrule
\end{tabular}
\end{table}

\begin{figure}[htbp]
  \centering
  \includegraphics[width=0.9\linewidth]{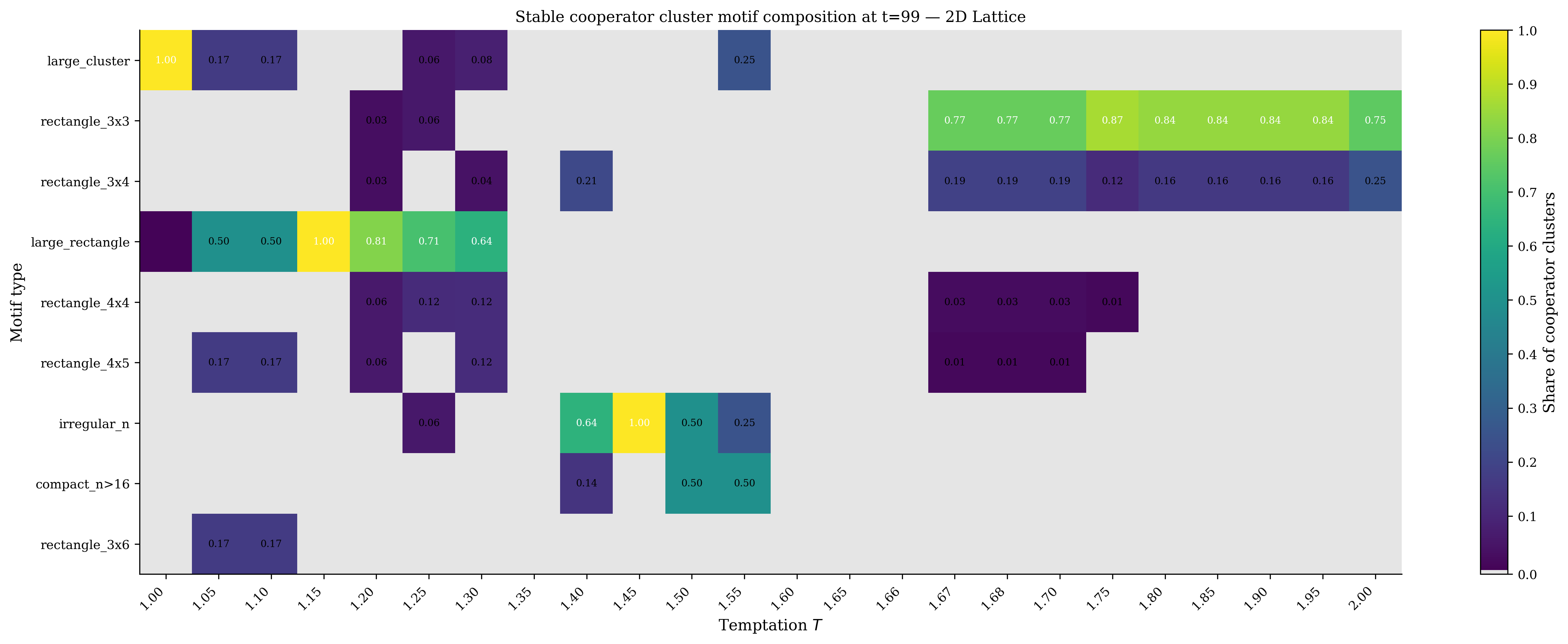}
  \caption{Stable cooperator cluster motif composition at \(t = 99\) for 2D lattices. Each cell shows the share of clusters of a given motif at temptation \(T\); light grey cells correspond to zero frequency.}
  \label{fig:stable-motifs-2d}
\end{figure}

\begin{figure}[htbp]
  \centering
  \includegraphics[width=0.9\linewidth]{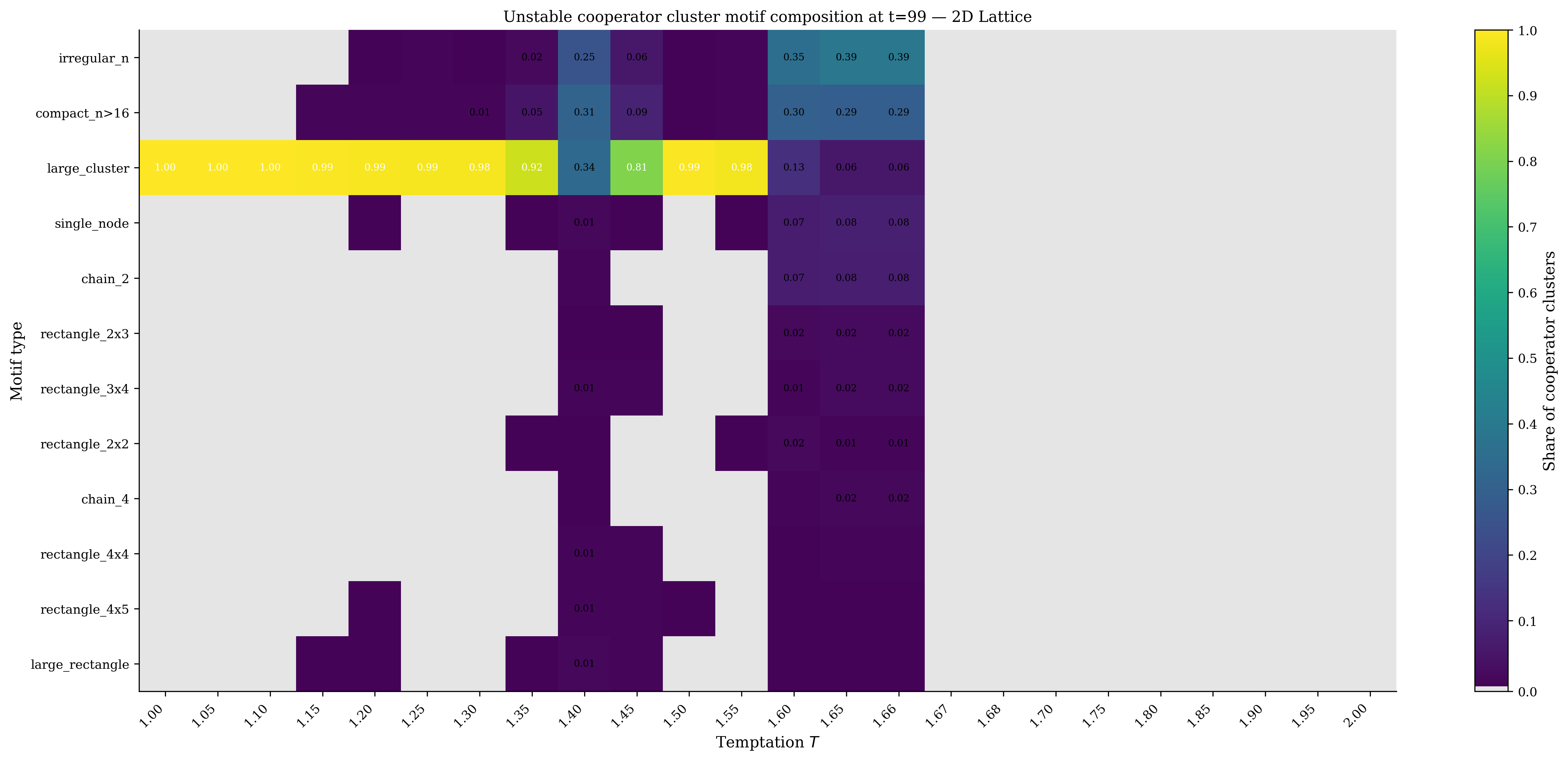}
  \caption{Unstable cooperator cluster motif composition at \(t = 99\) for 2D lattices.}
  \label{fig:unstable-motifs-2d}
\end{figure}

\begin{figure}[htbp]
  \centering
  \includegraphics[width=0.9\linewidth]{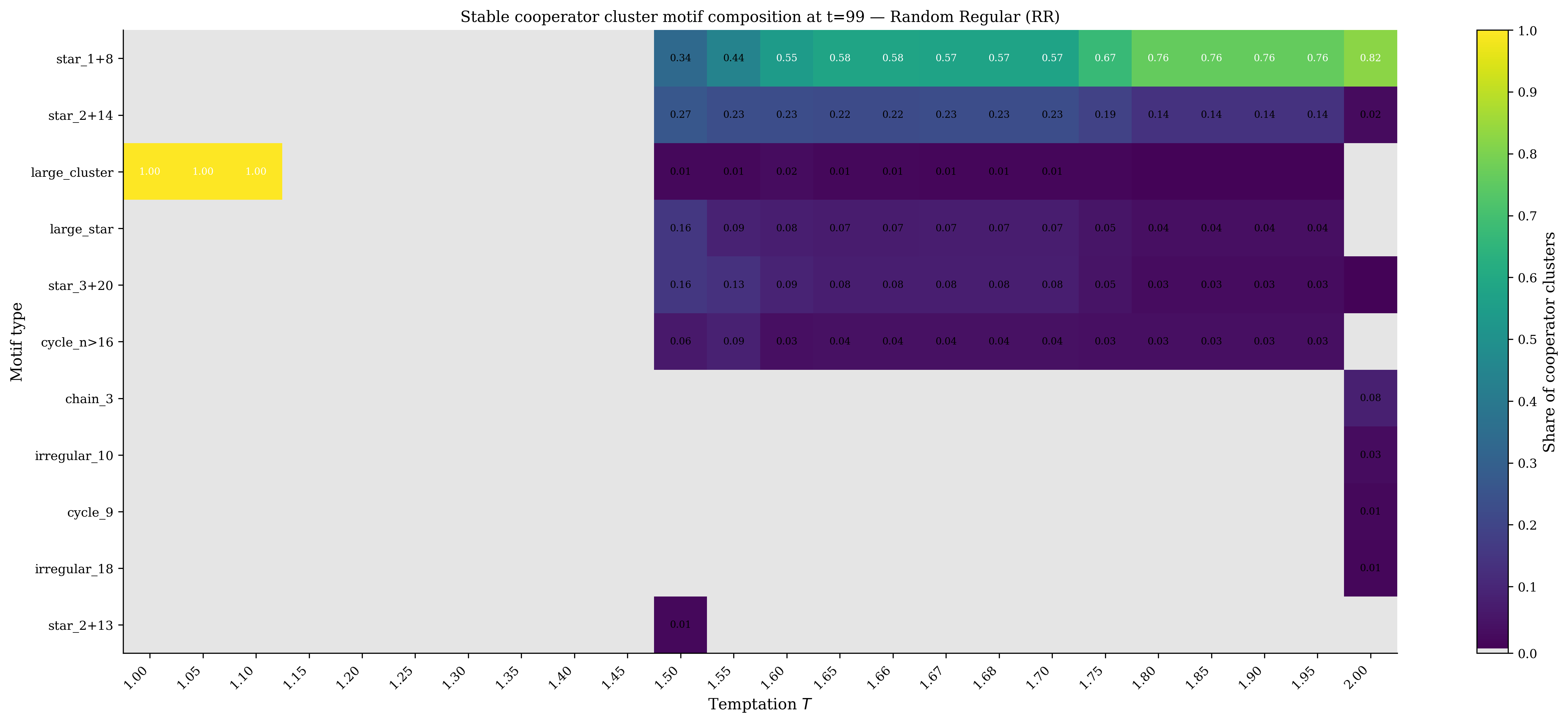}
  \caption{Stable cooperator cluster motif composition at \(t = 99\) for random-regular graphs.}
  \label{fig:stable-motifs-rr}
\end{figure}

\begin{figure}[htbp]
  \centering
  \includegraphics[width=0.9\linewidth]{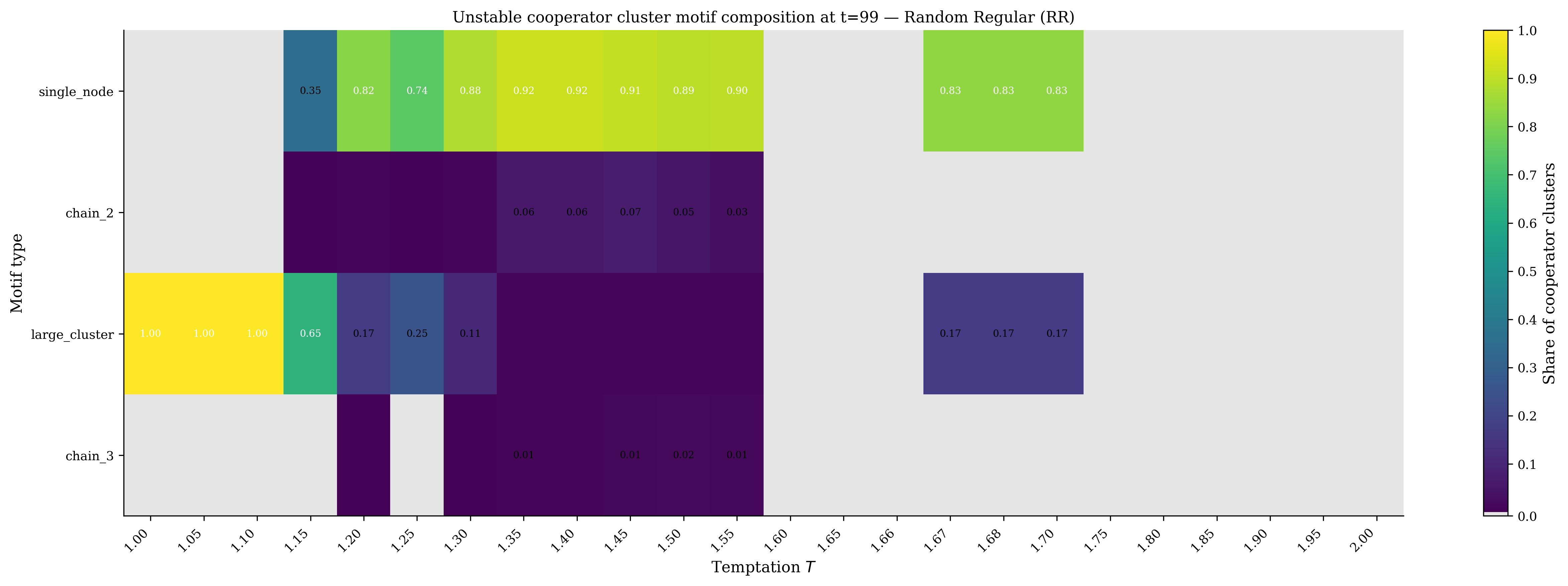}
  \caption{Unstable cooperator cluster motif composition at \(t = 99\) for random-regular graphs.}
  \label{fig:unstable-motifs-rr}
\end{figure}

\subsection{Random–regular network (\(k=8\))}
\label{sec:results:rr}

\subsubsection{Local payoff landscape of a \((1+m)\)-star}
\label{sec:results:rr:motif}

Because the RR topology contains no geometric blocks, the smallest
self–sustaining cooperative structure is a star.
Let a hub \(h\) (degree \(8\)) be a cooperator and let
\(m\) of its leaves (\(0\le m\le8\)) also be cooperators.
A leaf \(d\) that is still a defector has

\[
\pi_h = m, 
\qquad
\pi_d =
\begin{cases}
\;T     & \text{if no other cooperative neighbour},\\[4pt]
\;qT    & \text{if $q\ge1$ stray cooperators are adjacent to $d$.}
\end{cases}
\]

Under unconditional imitation \(d\to C\) iff \(\pi_h>\pi_d\).
Hence

\begin{equation}
\boxed{\;m>qT\;}.
\tag{2}
\end{equation}

Equation~(2) controls both star completion and erosion.  Two limiting cases are important:

\begin{itemize}
  \item \textbf{Isolated star (\(q=1\)).}  
        A hub with \(m\ge3\) converts every missing spoke as long as
        \(T< m\).  In particular, an eight–leaf star (size 9) is always
        stable for \(T<8\).
  \item \textbf{Single stray cooperator (\(q=2\)).}  
        If any stray \(C\) sits next to \(d\) the condition tightens to
        \(m>2T\).  For \(T\ge1.6\) even a hub with \(m=5\) cannot beat a
        boosted defector. Incomplete stars freeze and later disappear.
\end{itemize}

\subsubsection{Growth and arrest of star-like motifs}
\label{sec:results:rr:regimes}

Inequality (2) partitions the parameter space:

\begin{center}
\begin{tabular}{@{}lll@{}}
Condition on \(T\)              & Star dynamics                                         & Dominant clusters\\
\(T<1.10\)                     & any \((1+m)\)-star with \(m\ge3\) self–completes      & large merged domains\\
\(1.10\le T<1.50\)             & stars with \(m\le5\) stall if \(q=2\)                 & many unstable domains\\
\(1.50\le T\le2.00\)           & isolated star-like motifs survive                         & star-like motifs \\
\end{tabular}
\end{center}

The transition around \(T\approx1.5\) is therefore \emph{kinetic},
arising from the progressive failure of incomplete stars to finish
recruiting their leaves.

\begin{figure}[ht]
  \centering
  \includegraphics[width=.7\linewidth]{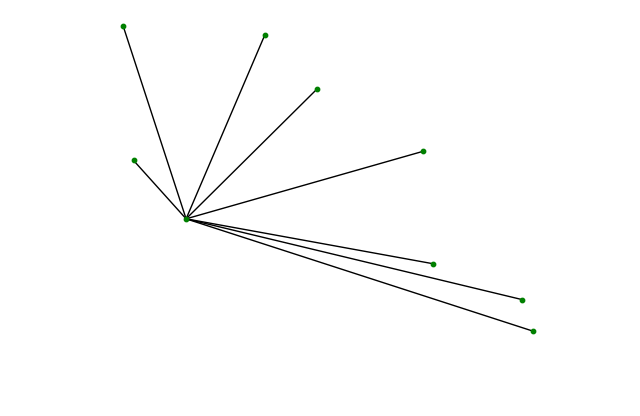}
  \caption{%
    \textbf{star\_1+8} star found on the degree-8 random–regular
    network for $T\gtrsim1.5$.  
    One hub (pay-off $8$) feeds eight leaves; each leaf therefore sees a
    neighbour with pay-off $8>T$ and never imitates an outside defector.
    The configuration is neutrally stable for every $T<8$.}
  \label{fig:rr_star_1+8}
\end{figure}

\begin{figure}[ht]
  \centering
  \includegraphics[width=.7\linewidth]{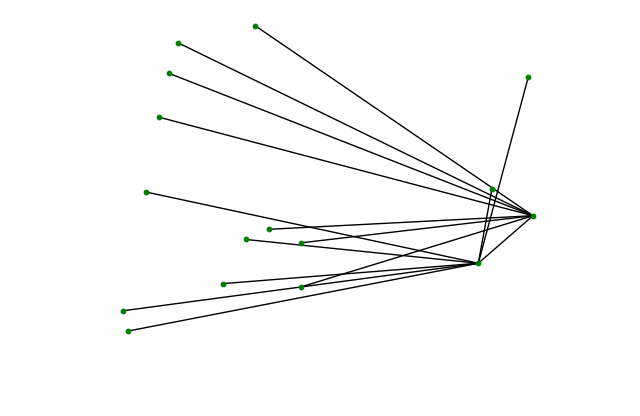}
  \caption{%
    \textbf{star\_2+14} star: two connected hubs (each pay-off $8$)
    jointly serve fourteen leaves, yielding the same stability condition
    as the 1\,+\,8 star.  Such twin-hub stars dominate the residual
    cooperation plateau for $T\gtrsim1.5$ when they manage to nucleate
    in the initial timesteps.}
  \label{fig:rr_star_2+14}
\end{figure}

Figures~\ref{fig:rr_star_1+8} and~\ref{fig:rr_star_2+14} display the
\texttt{star\_1+8} and \texttt{star\_2+14} tree-like motifs,
respectively, for visual reference.

\subsubsection{Numerical census and kinetic interpretation.}
\label{sec:results:rr:numerics}

Table~\ref{tab:rr-clusters} summarises the behaviour of the degree--8 random--regular network at $t=99$, reporting the global cooperation level $f_C$, the mean numbers of stable and unstable cooperator clusters $\langle N_s\rangle$ and $\langle N_u\rangle$, and the dominant stable motifs as a function of $T$. Figures~\ref{fig:stable-motifs-rr} and~\ref{fig:unstable-motifs-rr} complement this census by displaying the full motif composition of stable and unstable clusters at $t=99$ across temptation values. Together with the local star inequality~(2) in Section~\ref{sec:results:rr:motif}, these statistics support a kinetic picture in which cooperation is limited by the completion of a small number of star--like traps, rather than by a collective phase transition.

\begin{itemize}
\item[(a)]
  \textbf{$T \le 1.10$.}
For $T \in \{1.00,1.05,1.10\}$ the network almost always reaches a single macroscopic cooperative domain, with
$f_C \approx 0.91$--$0.92$, $\langle N_s\rangle \approx 0.73$, and $\langle N_u\rangle \approx 0.26$ (Table~\ref{tab:rr-clusters}).
The motif census shows that essentially all stable clusters are labelled \texttt{large\_cluster}, with share $1.000$ at these values of $T$.
This regime corresponds to the completion regime described in Section~\ref{sec:results:rr:regimes}, where any $(1{+}m)$--star with $m\ge 3$ quickly recruits its missing leaves and separate cooperative motifs tend to merge into larger components. See Fig.~\ref{fig:RR_1}.

\begin{figure}[ht]
  \centering
  \includegraphics[width=0.8\textwidth]{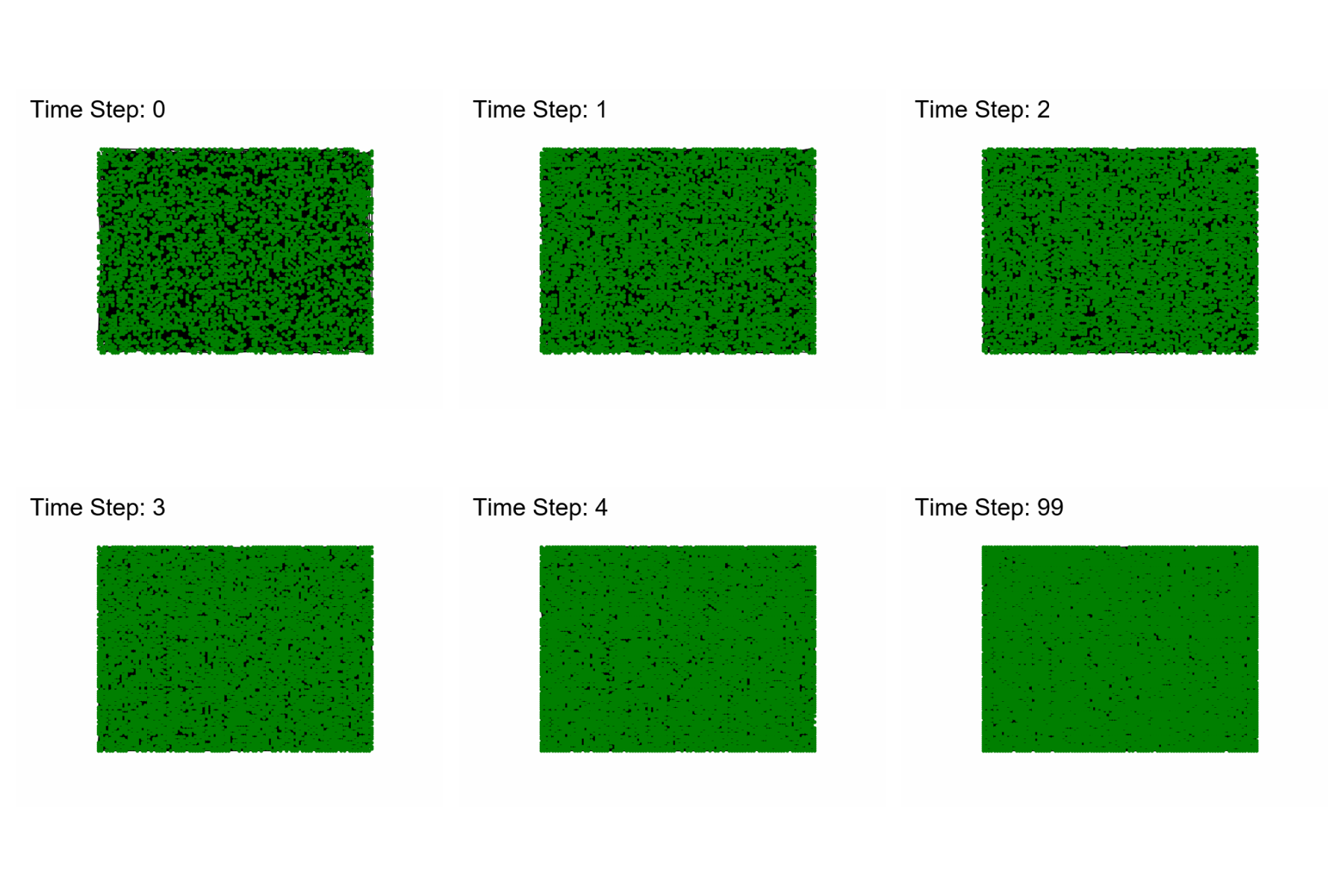}
  \caption{Time evolution of cooperator clusters on a $100\times100$ random-regular graph for $T=1.0$. Snapshots at time steps $t=0,1,2,3,4,99$ illustrate the expansion of cooperative bricks and other irregular shapes and the stability of certain defector structures at $t=99$.}
  \label{fig:RR_1}
\end{figure}

\item[(b)]
  \textbf{$1.10 < T < 1.50$.}
As $T$ increases beyond $1.10$, $f_C$ decays monotonically from $0.635$ at $T=1.15$ to $0.190$ at $T=1.45$, while $\langle N_s\rangle$ collapses to zero and $\langle N_u\rangle$ grows by two orders of magnitude, reaching $\langle N_u\rangle \approx 180$ at $T=1.45$ (Table~\ref{tab:rr-clusters}).
Figure~\ref{fig:unstable-motifs-rr} shows that these unstable components are dominated by isolated cooperators (\texttt{single\_node}), with a smaller contribution from remnants of \texttt{large\_cluster} components and short \texttt{chain\_2}/\texttt{chain\_3} fragments.
This is consistent with a fragmentation regime in which many cooperative seeds appear but fail to consolidate into persistent traps by $t=99$. Kinetically, this is the regime where boosted defectors begin to obstruct star completion.
When a missing spoke has additional cooperative neighbours, its payoff increases to $qT$ and the completion condition $m>qT$ becomes harder to satisfy.
In particular, for $q=2$ the requirement becomes $m>2T$, so as $T$ approaches $1.5$ incomplete stars with small $m$ (e.g., $m\le 3$) cannot finish recruiting their leaves, and larger $q$ makes completion even less likely.
Because the random--regular graph is locally tree--like, perturbations from
stray cooperators occur across many leaves, so partial stars are increasingly blocked rather than completed.
The result is many unstable fragments and a drop in global cooperation. See Fig.~\ref{fig:RR_2}.

\begin{figure}[ht]
  \centering
  \includegraphics[width=0.8\textwidth]{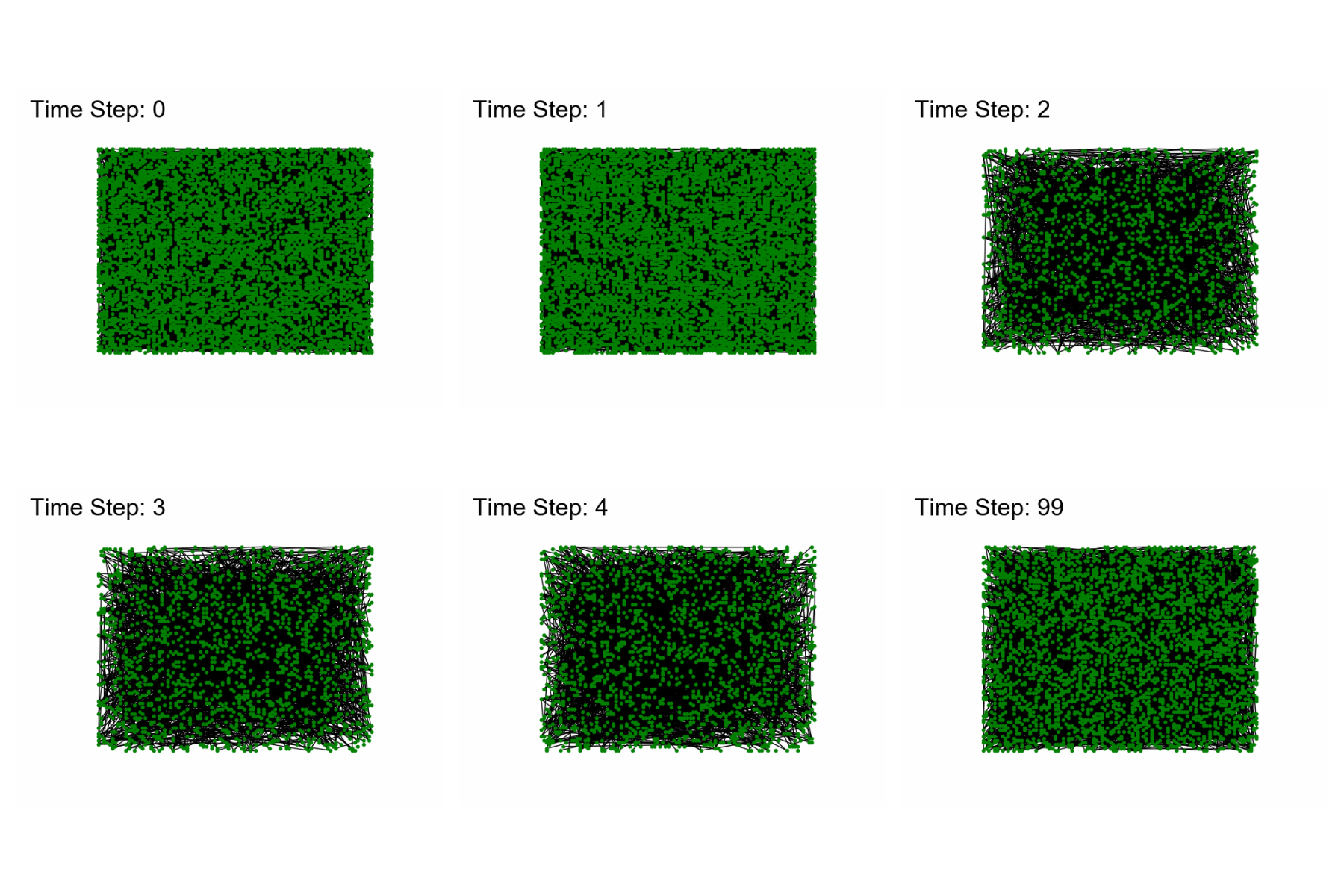}
  \caption{Time evolution of cooperator clusters on a $100\times100$ random-regular graph for $T=1.4$. Snapshots at time steps $t=0,1,2,3,4,99$ illustrate the expansion of cooperative bricks and other irregular shapes and the stability of certain defector structures at $t=99$.}
  \label{fig:RR_2}
\end{figure}

\item[(c)]
  \textbf{$T \ge 1.50$.}
Beyond $T \approx 1.5$ the system enters a rare--event plateau.
The average cooperation level drops from $f_C \approx 7\times 10^{-2}$ at $T=1.50$ to $f_C \sim 10^{-3}$ for $1.55 \le T \le 1.95$, and down to $f_C \approx 2.6\times 10^{-4}$ at $T=2.00$ (Table~\ref{tab:rr-clusters}). 
At the same time $\langle N_u\rangle$ falls back to essentially zero, while the number of stable clusters stabilises around $\langle N_s\rangle \approx 2$--$3$ for $1.60 \le T \le 1.80$.
The motif census in Table~\ref{tab:rr-clusters} and the stable--cluster heatmap (Fig.~\ref{fig:stable-motifs-rr}) show that these surviving components are almost exclusively \texttt{star\_1+8} and \texttt{star\_2+14} trees, with combined shares above $0.80$ for all $T \gtrsim 1.6$.
Other motifs (cycles, chains, irregular clusters) are either transient or statistically negligible at $t=99$, and only a small number of tiny \texttt{chain\_3} motifs appear as stable at $T=2.0$. See Fig.~\ref{fig:RR_3}.

\begin{figure}[ht]
  \centering
  \includegraphics[width=0.8\textwidth]{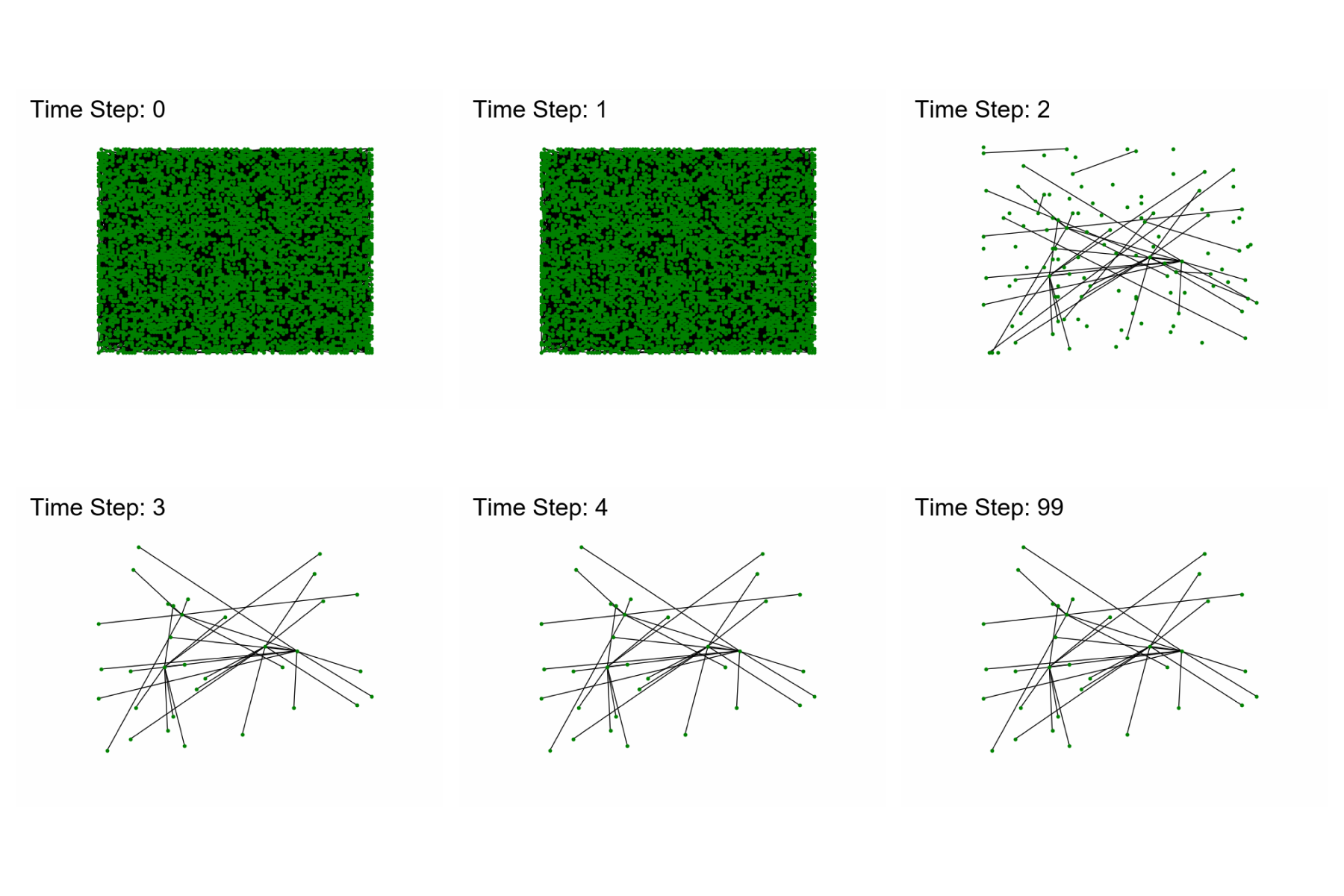}
  \caption{Time evolution of cooperator clusters on a $100\times100$ random-regular graph for $T=1.8$. Snapshots at time steps $t=0,1,2,3,4,99$ illustrate the expansion of cooperative bricks and other irregular shapes and the stability of certain defector structures at $t=99$.}
  \label{fig:RR_3}
\end{figure}

\end{itemize}

In summary, at high values of $T$, residual cooperation is explained by the early appearance of a few fully
completed stars. Completed stars are extremely rare under random initial
conditions because they require a perfectly coordinated local neighbourhood and
must avoid interference from nearby stray cooperators during the first time steps.
Crucially, once a star is complete, inequality~(2) implies neutral stability for
all $T<8$, so in our range $T\le 2$ the stability of completed stars is
essentially $T$--independent. The observed decay of $f_C(T)$ in
Table~\ref{tab:rr-clusters} is therefore not driven by an intrinsic loss of stability of completed stars, but by a kinetic
bottleneck. As $T$ increases, incomplete stars and other cooperative fragments
are progressively blocked by boosted defectors, fail to complete, and are
eventually eroded.

\section{Conclusion}

In this paper we revisited the spatial Prisoner’s Dilemma using exactly the
parameter set of the canonical Nowak and May study, so that our results can be
read as a direct extension of their seminal work \cite{Nowak1992}.

Across $R=1000$ runs on networks of size $N=10^4$ (with $t_{\max}=99$), using
degree $k=8$, payoffs $(R,S,T,P)=(1,0,T,0)$ with $T\in[1,2]$, and synchronous
unconditional imitation, we find two robust signatures at high $T$. First, on
the two-dimensional Moore lattice, for $T>5/3$ every connected cooperator component that
persists in an absorbing configuration is a fully occupied rectangle whose side lengths are at least $3$ (i.e., a $w\times h$ brick with
$w,h\ge 3$). Second, on random--regular graphs, for $T\gtrsim 1.5$ stable cooperator
components are dominated by fully completed stars (primarily \texttt{star\_1+8}
and \texttt{star\_2+14}), while other shapes are rare.

In both topologies studied here, the most frequent surviving motif at
high $T$ contains a cooperator whose entire neighbourhood is cooperative
($n=k=8$). This suggests a broader
hypothesis: at high $T$, under payoffs $(R,S,T,P)=(1,0,T,0)$ and synchronous unconditional
imitation on $k$--regular graphs, once the dynamics becomes nucleation limited,
the most frequent stable finite cooperator component is a motif that contains a
cooperator vertex with $n=k$ cooperative neighbours (equivalently, a cooperator
whose entire neighbourhood is cooperative).

Furthermore, in both topologies the cooperation drop is mainly kinetic. It is not driven by a
sudden loss of intrinsic viability of the most resilient motifs (bricks on the
lattice, completed stars on RR graphs), which remain stable in the relevant
parameter range. Instead, as $T$ increases the dynamics becomes increasingly
dominated by interference. That is, subcritical shapes stop expanding, and stray
cooperators boost nearby defectors, creating local noise that blocks the
completion of growing motifs. The effective density of successful seeds then
collapses, so cooperative domains fail to coalesce and the global cooperation
level drops.

Our results reinforce a central lesson from three decades of work on spatial
evolutionary games \cite{Szabo2007,Roca2009,Ohtsuki2006,Traulsen2006,Cuesta2015}, namely, network reciprocity is not one mechanism but a family of mechanisms whose
effectiveness depends on the update rule, the topology, and the relevant
time-scales. From an engineering viewpoint, a key implication is that
cooperation at high temptation is limited not only by payoffs but also by structural and physical constraints that determine which local
cooperative patterns can actually assemble and persist. This resonates with
recent work showing that physicality can strongly shape network structure and
function \cite{posfai2024impact,stopczynski2018physical,kerssies2024connect,segovia2020network}. More broadly, it remains an active research question how small initial shocks, homophily, and local cluster concentration shape complex
contagion and collective outcomes
\cite{mcpherson2001birds,watts2002simple, centola2010spread, segovia2024cross}. Two avenues for future research remain open: first, to derive analytical
results that confirm our numerical findings; and second, to test how far they
generalise to networks with different degrees.

\section*{Data Availability}

Electronic supplementary material and simulation code are available at \url{https://github.com/jsegoviamartin/Topological_traps_in_evolutionary_games} and \cite{segovia2025topological}.

\bibliographystyle{unsrtnat}
\bibliography{main}

\end{document}